\documentclass{nature}
\usepackage{graphicx}
\usepackage{color}
\usepackage{footmisc}
\usepackage{hyperref}

\def \msun{\ifmmode{{\rm\ M}_\odot}\else{${\rm\ M}_\odot$}\fi}
\def \zsun{\ifmmode{{\rm\ Z}_\odot}\else{${\rm\ Z}_\odot$}\fi}
\newcommand{\ha}{H$\alpha${}}
\newcommand{\hb}{H$\beta${}}
\newcommand{\hii}{H\,{\sc ii}{}}
\newcommand{\feii}{Fe\,{\sc ii{}}}
\newcommand{\oi}{O\,{\sc i{}}}
\newcommand{\oii}{O\,{\sc ii{}}}
\newcommand{\caii}{Ca\,{\sc ii{}}}
\newcommand\as{${''}$}
\newcommand{\kms}{kms$^{-1}$} 
\newcommand{\niii}{Ni\,{\sc ii{}}}
\newcommand\am{${'}$}
\def\iraf{{\sc iraf}}

\usepackage{newfloat}
\DeclareFloatingEnvironment[name={Supplementary Figure}]{suppfigure}
\DeclareFloatingEnvironment[name={Supplementary Table}]{supptable}
\title{The lowest metallicity type II supernova from the\\ highest mass red-supergiant progenitor}

\author{J.~P. Anderson,$^{1*}$ L. Dessart,$^{2}$, C.~P. Guti\'errez,$^{3}$ T. Kr\"uhler,$^4$
L. Galbany,$^{5}$ A. Jerkstrand,$^{6}$ S.~J. Smartt,$^{6}$ C. Contreras,$^7$ N. Morrell,$^7$
M.~M. Phillips,$^{7}$ M.~D. Stritzinger,$^8$ E.~Y. Hsiao,$^{9}$, S. Gonz\'alez-Gait\'an,$^{10,11}$
C. Agliozzo,$^{10,12}$ S. Castell\'on,$^9$ K.~C. Chambers,$^{13}$ T.~-W. Chen,$^4$ H. Flewelling,$^{13}$
C. Gonzalez,$^7$ G. Hosseinzadeh,$^{14,15}$ M. Huber,$^{13}$
M. Fraser,$^{16}$
C. Inserra,$^{3}$
E. Kankare,$^{6}$ S. Mattila,$^{17}$ E. Magnier,$^{13}$ K. Maguire,$^{6}$, T.~B. Lowe,$^{13}$
J. Sollerman,$^{18}$ M. Sullivan,$^{3}$ D.~R. Young,$^{6}$ S. Valenti$^{19}$}

\begin{document}

\maketitle

\begin{affiliations}
\item European Southern Observatory, Alonso de C\'ordova 3107, Casilla 19, Santiago, Chile
\item Unidad Mixta Internacional Franco-Chilena de Astronom\'ia (CNRS UMI 3386), Departamento de Astronom\'ia, Universidad de Chile, Camino El Observatorio 1515, Las Condes, Santiago, Chile
\item School of Physics and Astronomy, University of Southampton, Southampton, SO17 1BJ, UK
\item Max-Planck-Institut f{\"u}r extraterrestrische Physik, Giessenbachstra{\ss}e, D-85748 Garching, Germany
\item PITT PACC, Department of Physics and Astronomy, University of Pittsburgh, Pittsburgh, PA 15260, USA
\item Astrophysics Research Centre, School of Mathematics and Physics, Queens University Belfast, Belfast BT7 1NN, UK
\item Carnegie Observatories, Las Campanas Observatory, Casilla 601, La Serena, Chile
\item Department of Physics and Astronomy, Aarhus University, Ny Munkegade 120, 8000 Aarhus C, Denmark
\item Department of Physics, Florida State University, 77 Chieftan Way, Tallahassee, FL 32306, USA
\item Millennium Institute of Astrophysics, Universidad de Chile, Casilla 36-D, Santiago, Chile
\item Center for Mathematical Modelling, University of Chile, Beauchef 851, Santiago, Chile
\item Departamento de Ciencias Fisicas, Universidad Andres Bello, Avda. Republica 252, Santiago, Chile
\item Institute for Astronomy, University of Hawaii, 2680 Woodlawn Drive, Honolulu, HI 96822
\item Las Cumbres Observatory, 6740 Cortona Dr Ste 102, Goleta, CA 93117-5575, USA
\item Department of Physics, University of California, Santa Barbara, CA 93106-9530, USA
\item O'Brien Centre for Science, North University College Dublin, Belfield, Dublin 4, Ireland
\item Tuorla Observatory, Department of Physics and Astronomy, University of Turku, V\"ais\"al\"antie 20, 21500, Piikki\"o, Finland
\item Department of Astronomy and the Oskar Klein Centre, Stockholm University, AlbaNova, SE-106 91 Stockholm, Sweden
\item Department of Physics, University of California, Davis, CA 95616, USA

\end{affiliations}

\begin{abstract}
Red supergiants have been confirmed as the progenitor stars of the majority of hydrogen-rich
type II supernovae\cite{sma15}. However, while such stars are observed with masses 
$>$25\msun \cite{lev05}, detections of $>$18\msun\ progenitors remain elusive\cite{sma15}.
Red supergiants are also expected to form at all metallicities, but discoveries 
of explosions from low-metallicity progenitors are scarce.
Here, we report observations of the type II supernova, SN~2015bs, for
which we infer a progenitor metallicity of $\leq$0.1\zsun\ from comparison to photospheric-phase
spectral models\cite{des13},
and a Zero Age Main-Sequence mass of 17-25\msun\ through comparison to nebular-phase spectral models\cite{des10,jer12}.
SN~2015bs displays a normal `plateau' light-curve morphology, 
and typical spectral properties, implying a
red supergiant progenitor. 
This is the first example of such a high mass progenitor for a `normal' type II supernova,
suggesting a link between high mass red supergiant explosions and low-metallicity progenitors.
\end{abstract}

Type II supernovae (SNe~II) are the most abundant stellar explosions in 
the Universe,
as measured in volume-limited samples\cite{li11}. 
(We use `SNe~II' to refer to all objects showing flat or declining $V$-band light curves, together with broad \ha\ features, 
excluding type IIn, IIb and SN~1987A-like events.)
They are the only SN type with robust constraints
on their progenitor stars\cite{sma15}, providing 
direct evidence for red supergiant (RSG) progenitors and confirming
results from light-curve modelling\cite{fal77}. Pre-explosion images constrain
their initial mass to be 8.5-18\msun \cite{sma15}. The lack of
progenitors above this mass is referred to as the `red supergiant problem'\cite{sma09}, 
given that at least some stars $>$18\msun\ should be viable SN~II progenitors\cite{woo02},
with the exact mass limit being dependent on rotation, 
metallicity and mass-loss\cite{hir04,chi13}.
This is also seen when comparing nebular-phase spectra ($>$200 days 
post explosion, +200\,d)
with SN~II explosion models\cite{des10,jer14,jer12,jer15,sil16}.
A number of solutions to this issue have been proposed. 
\cite{wal12} suggested that the inclusion of unaccounted for circumstellar dust around progenitors
could translate to higher luminosities and therefore higher masses.
It has been argued that the problem disappears if accurate bolometric corrections
are used to estimate progenitor luminosities\cite{dav18}. The predicted upper 
mass limit for producing SNe~II decreases in rotating models\cite{hir04} and when employing
higher RSG mass-loss rates\cite{chi13}. This opens the possibility that progenitors above 20\msun\ may not explode as SNe~II, but
as SNe~IIb or SNe~Ib.
However, it has also been argued\cite{sma15} that this dearth of massive progenitors 
is due to RSGs collapsing to a black holes 
with no 
(or a weak/faint)
accompanying SN. This latter scenario is supported by the observed bimodal 
distribution of compact remnants\cite{koc14}, and the recent detection of a vanishing 25\msun\ RSG star\cite{ada17}.\\
\indent Historical SN surveys prioritised SN detection over completeness concentrating on
observations of bright, nearby galaxies, where the majority of the star formation (SF) takes place at solar metallicity. 
This led to a lack of SNe found in low-luminosity, low-metallicity galaxies. While
modern surveys are rectifying this situation\cite{arc10}, 
samples of SNe~II in hosts of low metallicity ($\leq$0.5\zsun)
are still lacking\cite{sto13,des14,and16}. We therefore started a follow-up program to study SNe~II discovered in galaxies dimmer than
--18.5 in the $B$-band, through the \textit{Public ESO Spectroscopic Survey of Transient Objects} (PESSTO)\cite{sma15b}.\\
\indent 
On the 25th of September 2014, the \textit{Catalina Real-Time Transient Survey} (CRTS)\cite{dra09} discovered the apparently
host-less SN CSS140925:223344-062208. 
It was also recovered by the CRTS in the Mount Lemmon facility, 
and detected by the \textit{Panoramic Survey Telescope and Rapid Response System} 
(Pan-STARRS1\cite{hub15}: https://star.pst.qub.ac.uk/ps1threepi/psdb/, 
hereafter the SN is designated as the IAU confirmed name of SN~2015bs).
A pre-SN non detection constrains its explosion epoch to be the 20th of 
September $\pm$5 days. The classification spectrum 
revealed Balmer lines on top of a blue continuum, indicative 
of a young SN~II. 
A redshift of around 0.02 was estimated from the SN spectrum.
Three additional optical spectra were obtained during 
the plateau phase, together with $B,V,r,i$ photometry. A year post explosion we also obtained integral
field spectroscopy of SN~2015bs and its surroundings.\\
\indent 
SN~2015bs displays a relatively luminous, but normal optical light-curve (Fig.\,1, and Supplementary Information, SI). 
At $\sim$50\,d, the spectrum of SN~2015bs is
dominated by the typical hydrogen Balmer lines observed in SNe~II (Fig.\,2a). However,
metal absorption lines are much less prominent in comparison to other SNe~II.
Spectral models produced from progenitors of 
different metallicity\cite{des14,des13} show that as metallicity decreases metal-line pseudo-equivalent
widths become weaker. Further, SNe~II occurring within 
lower-metallicity
\hii\ regions display weaker \feii\ 5018\,\AA\ lines\cite{and16}.
Fig.\,3 shows how the +57\,d spectrum of SN~2015bs is well matched by a model at 0.1\zsun, 
in contrast with SN~2012aw, whose strong metal lines support a super-solar metallicity progenitor.
Measuring the pseudo equivalent width of the \feii\ 5018\,\AA\ line,
we find 4.25$\pm$0.54\,\AA\ for SN~2015bs, and 3.61$\pm$1.29\,\AA\ for the 0.1\zsun\ model (11.33$\pm$0.71\,\AA\ is measured for 
the 0.4\zsun\ model), 
in support of a low-metallicity progenitor.\\
\indent Using our late-time spectroscopy, 
we identify the host of SN~2015bs at an angular
separation of 3.4\as\ from the SN (see SI Fig.\,1) 
that shows narrow \ha\ (from ionised gas within the galaxy) at a redshift
of 0.027, consistent with the spectra of SN~2015bs. 
We measure an absolute $r$-band magnitude for the host of --12.2 mag. 
This makes SN~2015bs the lowest-luminosity host for a
SN~II, being more than a magnitude fainter than the previously dimmest host\cite{sto13,arc10,tad16}.
Using well known galaxy luminosity--metallicity relationships
this translates to a host metallicity of 0.04\zsun$\pm$0.05\cite{arc10,tre04}.\\
\indent In addition to being the lowest metallicity SN~II 
studied to-date (as compared to all previous published SN~II environment metallicities\cite{sto13,tad16}),
SN~2015bs
is unique in its nebular phase. 
It presents striking differences compared to other SNe~II (Fig.\,2b). 
Dominant spectral lines at these epochs are [\oi] 6300,6364 \AA, \ha, and [\caii] 7291,7323 \AA.
In SN~2015bs [\oi] is as strong as \ha\ and [\caii]: in most other SNe~II \ha\ is stronger than either line, and 
[\caii] is stronger than
[\oi]. In addition, the nebular hydrogen line of SN~2015bs 
(Supplementary Fig.\,8) is broader than seen in other SNe~II.
Observations at nebular epochs can be used to constrain the properties of the helium core. Following the
the tight relation between helium-core mass and ZAMS\cite{woo95} (that is largely insensitive to metallicity up to $\sim$30\msun\cite{woo02,des13}), 
we thus constrain the progenitor mass of
SN~2015bs. One caveat is the way convection is treated in 1-D models, and the
associated uncertainties\cite{arn11} that may complicate the exact mapping to ZAMS mass.\\
\indent The absolute strength of [\oi] is an indicator of the helium core mass,
and nebular modelling of SNe~II reveals that as progenitor mass increases so does the strength of
[\oi] as compared to \ha\ and [\caii]\cite{jer12,fra87}. Our observations 
therefore suggest that SN~2015bs was the explosion of a higher mass progenitor
than previously observed SN~II. In the Supplementary Information (SI) 
we make comparisons between the +413\,d spectrum of SN~2015bs and 
spectra from 15 and 25\msun\ ZAMS models\cite{jer14} (Supplementary Fig.\,11). SN~2015bs displays significantly 
stronger [\oi] than the 15\msun\ model, suggesting a
higher mass progenitor than previous nebular-spectroscopic constraints.\\
\indent We make quantitative comparisons between SN~2015bs, our comparison SN~II sample, 
and models, using the
percentage of the [\oi] flux with respect to the total optical flux contained 
within the wavelength range of the nebular SN~2015bs spectrum (see Table 1). This is an 
alternative to using the luminosity of [\oi] normalised to the $^{56}$Co decay power. 
The $^{56}$Co-normalisation method has been used\cite{jer12} because
gamma-ray trapping also depends on ZAMS mass, errors from extinction
are moderated as the luminosity of [\oi] and the $^{56}$Ni mass estimates are affected similarly, and contamination
by background continuum is removed. Using the optical ratio
removes uncertainties associated to bolometric corrections used to estimate $^{56}$Ni masses.
SN~2015bs has a value of 15.4$\pm$0.7\%, which is 
at least twice higher than previously observed. This provides
further evidence that SN~2015bs arose from the highest mass SN~II progenitor to date.
SN~2015bs is closer to the percentage of the 19\msun\ model than that of 15\msun\ (Table 1), and we here constrain its
progenitor mass to be 17-18\msun.
Such a mass constraint lies at the upper limit of the mass range from direct progenitor detections -- while being larger than any 
previous nebular-spectrum constraints. However,
it is clear from Fig.\,2b that the nebular spectrum of SN~2015bs is significantly distinct from other SNe~II.
There is therefore a real difference in helium-core mass (and therefore progenitor mass) between SN~2015bs and previously studied
RSG explosions.\\
\indent One should note
that model line fluxes start to saturate above 19\msun\ due to line absorption in the increasingly
dense cores (see the relatively small increase in the [\oi] percentage for the 25\msun\ model). 
This means that models in the 20-30\msun\ range are only 20-30\%\ brighter in [\oi] than measured values, 
and cannot be ruled out considering model uncertainties. 
At the same time, model tracks at 20-25\msun\ are still over a factor 
of 3-6 brighter than 12-15\msun\ models, outlining the diagnostic power of using [\oi]
to determine progenitor mass. No previous SN~II nebular spectrum was consistent with models 
of ZAMS of much more than 15\msun, whereas -- within measurement and model uncertainties -- SN~2015bs is consistent with models of 19\msun\ and above.\\ 
\indent The broad nebular \ha\ emission of SN~2015bs can also be explained through the explosion of a
star with a higher helium-core mass. In SNe~II, the width of the nebular lines
reflect the velocity of the outer edge of the helium core, or equivalently the inner edge of the hydrogen-rich envelope\cite{des10}. 
Since the width of optical lines in SN~2015bs during the photospheric phase suggests a standard explosion energy,
this implies a larger fractional core mass, i.e. the helium-core material represents a larger 
fraction of the total mass and its outer edge is closer to the maximum velocity in the ejecta\cite{des10}. 
The high \ha\ nebular velocity of SN~2015bs (seen in Supplementary Figs\,7 and 8), therefore provides further evidence that SN~2015bs had a 
massive helium core. SN~2015bs has a Half-Width at Half Maximum (HWHM) nebular \ha\ velocity of
2127\,$\pm$\,308\,\kms, while inferred photospheric velocity at +50\,d is 5359\,$\pm$\,392\,\kms.
Making direct comparison to the hydrodynamic models of \cite{des10} (specifically velocities in their table 2), constrains
the progenitor mass of SN~2015bs to be between 20 and 25\msun.
While the velocity of the \ha\ nebular line is significantly higher for SN~2015bs than for any other SN~II, the velocities
of [\oi] are similar between SN~2015bs and the comparison sample (see Supplementary Fig.\,7). 
In low-mass SNe~II (ZAMS $\leq$12\msun), as much as $\sim$50\%\ 
of the nebular line emission of [\oi] and the majority of [\caii]
arises from the hydrogen-rich envelope, with the rest coming from the core\cite{jer12,mag12}. 
However, assuming a higher mass for SN~2015bs, [\oi] emission becomes dominated
by oxygen in the core rather than primordial oxygen in the envelope. This naturally explains the 
relatively high ratio of hydrogen to oxygen velocities as compared to other SNe~II (dominated by lower core-mass events),
and gives further support for a 20-25\msun\ progenitor for SN~2015bs. We note that such a difference between \ha\ and [\oi] line velocities
was also observed in SN~1987A\cite{mag12}, whose progenitor was also of relatively low metallicity and high mass.\\
\indent The association of SN~2015bs with a 17-25\msun\ progenitor star at 0.1\zsun\ has
important implication for massive star evolution and explosions. 
Firstly, it shows that stars more
massive than 17\msun\ can explode, and that not all such massive progenitors
proceed to direct black-hole formation without any accompanying bright transient. Together with the 
recent identification of a vanishing 25\msun\ RSG star\cite{ada17}, this supports
the notion that there may be `islands of explodability' for massive stars\cite{oco11}: 
the generally greater mass accretion rate onto the proto-neutron star 
forming in higher mass stars may not systematically lead to a  
failed explosion\cite{mue12}. Secondly, the link between a high-mass RSG explosion and a low-metallicity 
progenitor opens the possibility that progenitors $>$20\msun\ can successfully explode as SNe~II \textit{if}
the metallicity is sufficiently low (mass-loss is lower), while at solar
metallicities the majority of such RSGs 
may lose enough mass to explode as SNe~IIb or Ib
(although the detection of a disappearing high-mass RSG at solar metallicity provides an obvious counter example that this
is not always the case).
A detection of a high-mass and metallicity progenitor for such SNe would provide confirmation of this possibility. 
These different interpretations are discussed further in the SI.
\\
\indent 
We have presented observations of SN~2015bs, a type II SN that exploded in the lowest-luminosity host galaxy for any SN~II discovered 
to date\cite{sto13,arc10,tad16}.
The weakness of metal lines in the photospheric-phase spectrum is consistent with models of SNe~II at low metallicity, and confirms
the utility of SNe~II as metallicity indicators\cite{des14,and16}.
The nebular spectrum is notably distinct, implying a more massive progenitor than all previously known SNe~II. 
The effects of sub-Small Magellanic Cloud metallicities ($<$0.4\zsun)
on SNe~II and massive star evolution are relatively unconstrained observationally.
The unique characteristics of SN~2015bs highlights the bias in the
current sample of SNe~II, with most events studied at around solar metallicity.
Current and future surveys will
broaden the SN~II parameter space, and further our knowledge
of the evolution and explosion of massive stars.

%% Put the bibliography here, most people will use BiBTeX in
%% which case the environment below should be replaced with
%% the \bibliography{} command.

\newpage
\begin{centering}
\textbf{\large{References}}
\end{centering}

\begin{addendum}
\item TK and TWC acknowledges support through the Sofja
Kovalevskaja Award to P. Schady from the Alexander von Humboldt
Foundation of Germany. AJ acknowledges funding by the European Union’s
Framework Programme for Research and Innovation Horizon
2020 under Marie Sklodowska-Curie grant agreement No
702538.
SJS acknowledges funding from the European Research Council under the European Union's Seventh Framework
Programme (FP7/2007-2013)/ERC Grant agreement no [291222] and STFC grants ST/I001123/1 and ST/L000709/1.
MDS, CC and EH gratefully acknowledge the generous support provided by the Danish Agency 
 for Science and Technology and Innovation  
realized through a Sapere Aude Level 2 grant.
MDS acknowledges funding by a research grant (13261) from the VILLUM FONDEN.
Support for SG is provided by the Ministry of Economy, Development, 
 and Tourism's Millennium Science Initiative through grant IC120009 awarded to 
 The Millennium Institute of Astrophysics (MAS), and CONICYT through 
 FONDECYT grant 3140566. 
Support for CA is provided by the Ministry of Economy, Development, 
 and Tourism's Millennium Science Initiative through grant IC120009 awarded to 
 The Millennium Institute of Astrophysics (MAS), and CONICYT through 
 FONDECYT grant 3150463. 
MF acknowledges the support of a Royal Society - Science Foundation Ireland University Research Fellowship.
KM acknowledges support from the STFC through an Ernest Rutherford Fellowship.
MS acknowledges support from EU/FP7-ERC grant 615929.
The work of the CSP-II has been supported by the National Science 
 Foundation under grants AST0306969, AST0607438, AST1008343, and AST1613426.
This work is based (in part) on observations collected at the European Organisation
for Astronomical Research in the Southern Hemisphere, Chile as part of PESSTO, (the Public ESO Spectroscopic
Survey for Transient Objects) ESO program 188.D-3003, 191.D-0935.
This work is based (in part) on observations collected at the European Organisation
for Astronomical Research in the Southern Hemisphere under ESO programme 296.D-5003(A).
This work was partly supported by the European Union FP7 programme through ERC grant number 320360.
Pan-STARRS is supported by NASA grants NNX08AR22G, NNX14AM74G.  
PS1 surveys acknowledge the PS1SC: University of Hawaii, MPIA Heidelberg, MPE Garching, 
Johns Hopkins University, Durham University, University of Edinburgh, Queen’s University Belfast, 
Harvard-Smithsonian CfA, LCOGT, NCU Taiwan, STScI, University of Maryland, Eotvos Lorand University, 
Los Alamos National Laboratory, and NSF grant No. AST-1238877.
Avishay Gal-Yam, Melina Bersten, Francisco F\"orster, John Hillier, Francesco Taddia and Claus Fransson are thanked
for useful discussions.
This research
has made use of: the NASA/IPAC Extragalactic Database (NED) 
which is operated by the Jet 
Propulsion Laboratory, California
Institute of Technology, under contract with the National Aeronautics;
\iraf, which is distributed 
by the National Optical Astronomy Observatory, which is operated by the 
Association of Universities for Research in Astronomy (AURA) under 
cooperative agreement with the National Science Foundation;
QfitsView; and the SDSS,
funding for the SDSS and SDSS-II has been provided by the Alfred P. Sloan Foundation, the Participating Institutions, the National Science Foundation, the U.S. Department of Energy, the National Aeronautics and Space Administration, the Japanese Monbukagakusho, the Max Planck Society, and the Higher Education Funding Council for England. The SDSS Web Site is http://www.sdss.org/.

 \item[Author contributions] JPA performed the analysis and wrote the manuscript. LD helped write
 the manuscript and provided comments on the physical interpretation. CPG provided
 specific measurements of pEWs of spectral lines and was part of the overall project to obtain these data. TK reduced
 the MUSE dataset. LG helped obtain the MUSE dataset. AJ provided comments on the physical interpretation of
 the nebular spectral comparisons. SJS is PI of the PESSTO project, through which spectra were obtained. CC provided
 calibrated photometry from the CSP-II. NM obtained the photometry from the CSP-II. MMP is PI of the CSP, which provided
 photometric data. MS is co-I on the CSP, which provided photometric data. EYH is co-I on the CSP, which provided photometric data.
 SGG analysed the light-curve data of SN~2015bs. CA was part of the PESSTO project, through which spectra were obtained. 
 SC obtained the photometry from the CSP-II. KCC provided photometry through the Pan-STARRS project. TWC  was part of the PESSTO project, through which spectra were obtained. 
 CG obtained the photometry from the CSP-II. GH provided a spectrum from LCO. MH provided photometry through the Pan-STARRS project.
 MF was part of the PESSTO project, through which spectra were obtained. CI was part of the PESSTO project, through which spectra were obtained.
 EK was part of the PESSTO project, through which spectra were obtained. SM was part of the PESSTO project, through which spectra were obtained.
 EM provided photometry through the Pan-STARRS project. KM was part of the PESSTO project, through which spectra were obtained. TBL provided photometry through the Pan-STARRS project.
 JS was part of the PESSTO project, through which spectra were obtained. MS was part of the PESSTO project, through which spectra were obtained. DY 
 was part of the PESSTO project, through which spectra were obtained. SV was part of the PESSTO project, through which spectra were obtained.

 \item[Competing Interests] The authors declare that they have no
competing financial interests.
 \item[Correspondence] Correspondence and requests for materials
should be addressed to J.~P. Anderson~(email: janderso@eso.org).
\end{addendum}

\newpage

\begin{table*}
\centering
\footnotesize
\begin{tabular}[t]{ccc}
\hline
SN  & Epoch (days post explosion) & [\oi] percentage (error) \\
\hline
2015bs & 413 & 15.4 (0.7) \\
\hline
1999em & 391 & 5.2 (1.0)  \\
2004et & 401 & 5.7 (0.9)  \\
2007aa & 376 & 6.1 (0.7)  \\
2009N  & 406 & 3.0 (1.0)  \\
2012A  & 393 & 6.6 (0.4)  \\
2012aw & 451 & 8.0 (0.9)  \\
2013ej & 388 & 8.7 (0.8)  \\
\hline
12\msun\  & 400 & 4.1 (0.4)  \\
15\msun\  & 400 & 8.6 (0.6)  \\
19\msun\  & 400 & 17.9 (0.8)  \\
25\msun\  & 400 & 19.6 (1.0) \\
\hline
1987A  & 398 & 9.1 (0.3)  \\
\hline	                                 
\hline      
\end{tabular}
\caption{Measured [\oi] 6300,6364 \AA\ fluxes for SN~2015bs and our comparison SN~II sample as a percentage of total `optical' flux (4800 to 9300\,\AA). 
In the first column we list the SN name, followed
by the epoch of the nebular spectrum (days post explosion) in column 2.
In column 3 we present the [\oi] 6300,6364 \AA\ flux as a percentage of the total `optical' flux. 
Note, for the 19\msun\ model this value is calculated by interpolating
between nebular model spectra at +369 and +451\,d. We also include 
values for SN~1987A for comparison.
Errors on percentages are derived from
the standard deviation of multiple flux measurements while making slight changes to the defined continuum level.
While our model comparison suggests a ZAMS mass between 17-18\msun, the [\oi] percentage for the 25\msun\ model is not 
significantly higher than the 19\msun\ model. In addition, SN~2015bs shows a much larger value than SN~1987A, for which a 18-20\msun\ progenitor has been 
invoked.}
\label{omuse}
\end{table*}

\begin{figure}
\includegraphics[width=17cm]{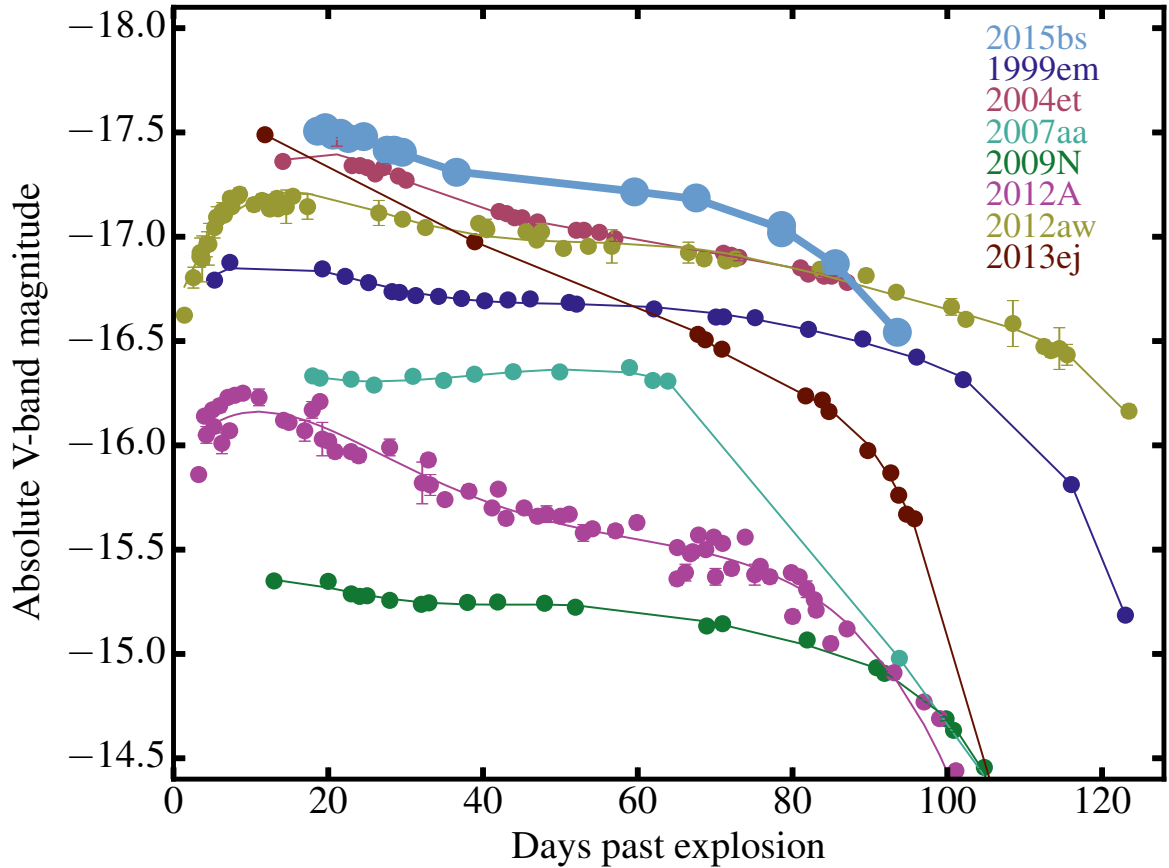}
\caption{Absolute $V$-band light-curves of SN~2015bs together with a comparison sample from the literature.
The light curve of SN~2015bs is shown in light blue.
Errors on the photometry of SN~2015bs are the propagated errors from the photometric calibration (those for the comparison
sample are taken from the literature).
While SN~2015bs falls on the bright side of the distribution, overall it displays a
normal light-curve morphology for a SN~II. The decline rate during the `plateau' phase appears
typical of SNe~II, as does the length of the optically thick phase duration. }
\label{abslcs}
\end{figure}

\begin{figure*}
\includegraphics[width=8.5cm]{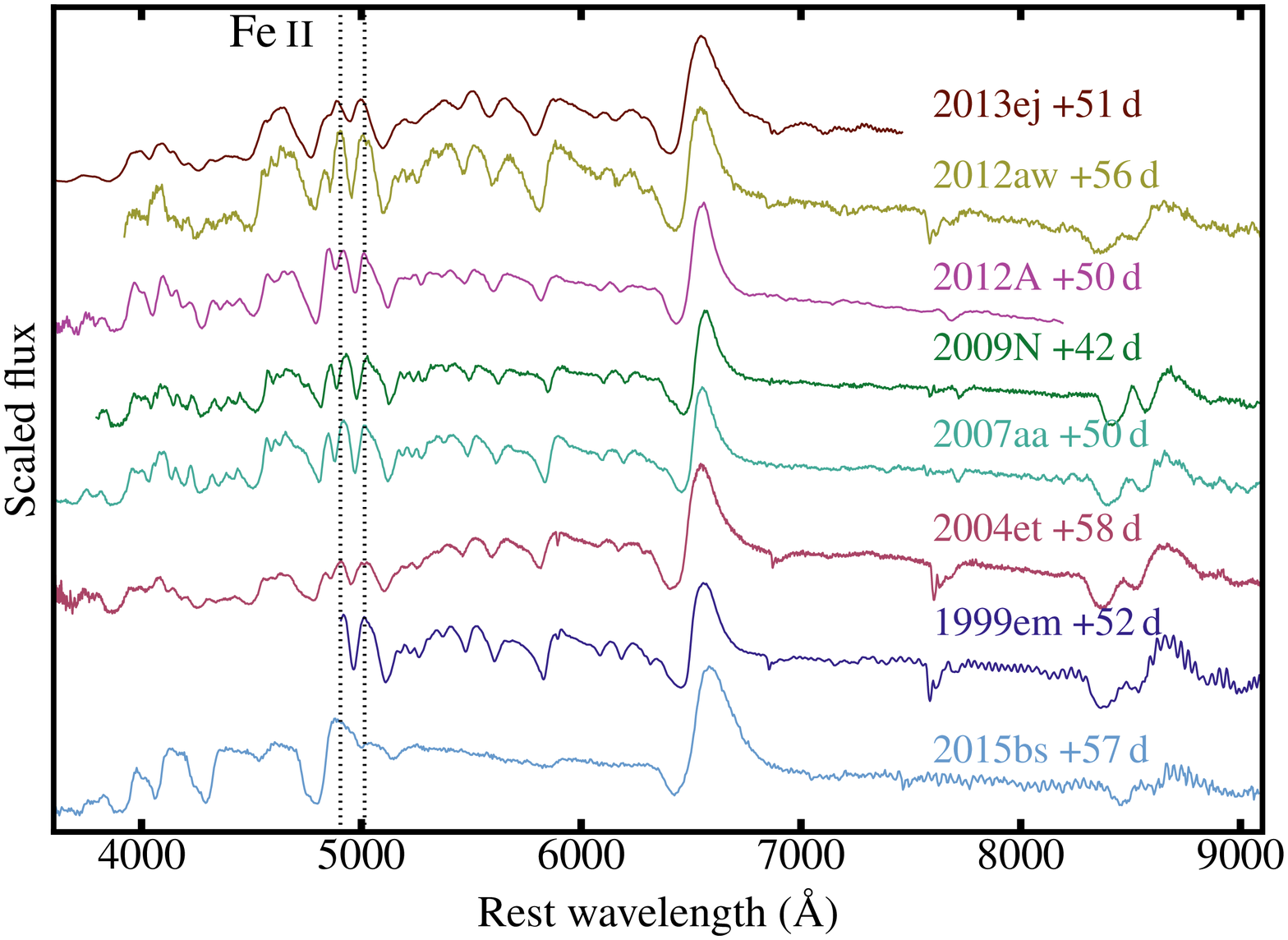}
\includegraphics[width=8.5cm]{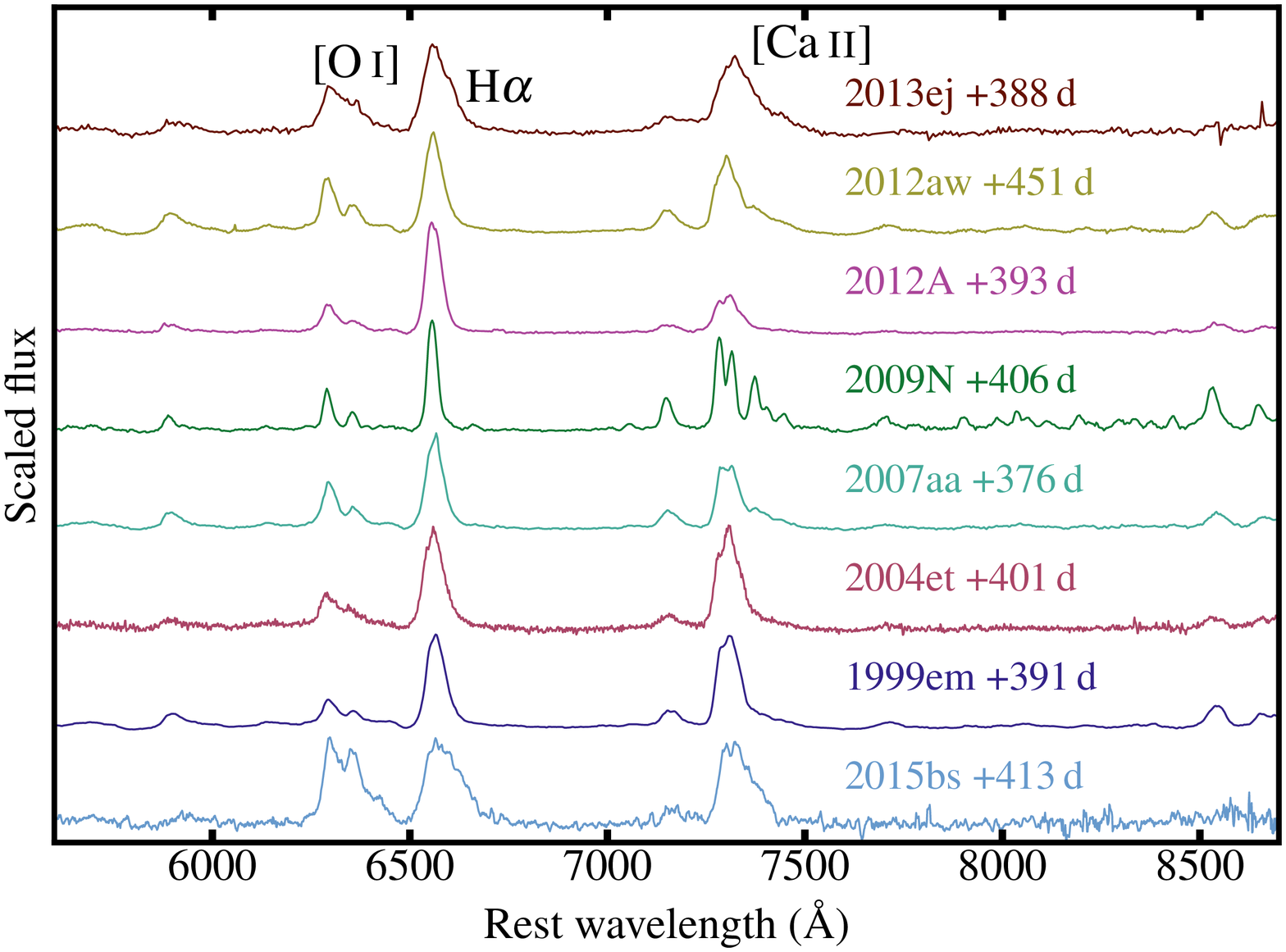}
\caption{Comparison of optical-wavelength spectra of SN~2015bs with literature SNe~II. \textit{a):} photospheric phase spectra. 
While the Balmer lines appear similar between SN~2015bs
and the comparison SNe, there is a clear lack of spectral
features in the blue part of the spectrum. 
The position of the \feii\ 5018\,\AA\ absorption line is indicated, bracketed 
by dotted black vertical lines. This line has been used as a proxy for progenitor metallicity\cite{and16},
and is significantly weaker in SN~2015bs.
\textit{b):} nebular phase spectra. The most prominent nebular lines are indicated on the spectrum of SN~2013ej:
[\oi] 6300,6364 \AA; \ha; and [\caii] 7291,7323 \AA.
While the comparison SNe~II all look quite similar -- apart from small changes in the width of emission lines -- SN~2015bs
is clearly distinct. The relative strength of oxygen is much higher, and in particular \ha\ is broader.}
\label{neb50comp}
\end{figure*}

\begin{figure}
\includegraphics[width=16cm]{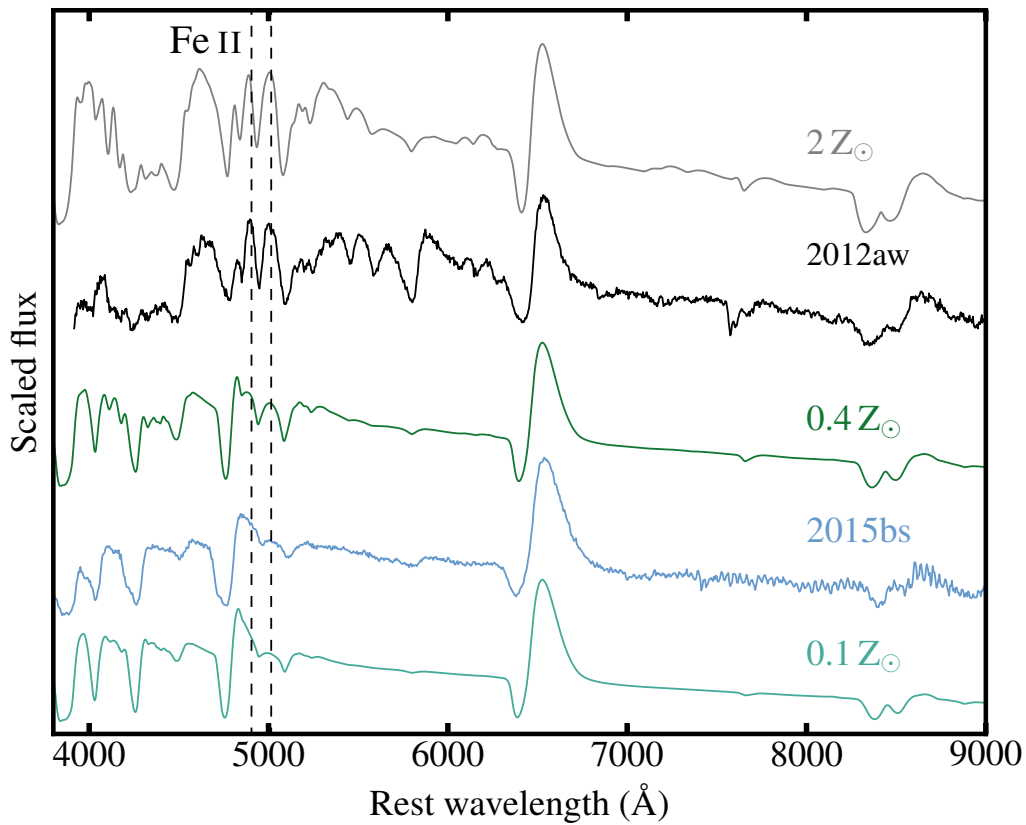}
\caption{Comparison of the 57\,d spectrum of SN~2015bs with 0.1\zsun\ and 0.4\zsun\ models at +50\,d\cite{des13}. 
We also present
an example SN~II from our comparison sample (SN~2012aw) which shows much more prominent metal lines, 
more consistent with the 2\zsun\ model.
It is clear that the 0.1\zsun\ model best matches
the spectrum of SN~2015bs (observe the regions bluer than 5000 \AA, and
specifically the strength of the \feii\ 5018\,\AA\ line, which is
bracketed by dashed lines), providing strong evidence for the low-metallicity nature of SN~2015bs.}
\label{01model}
\end{figure}

\clearpage 

\vspace{0.5cm}
\begin{centering}
\title{Methods}\\
\end{centering}

\begin{centering}
\textbf{1. CSS140925:223344-062208, aka SN~2015bs}\\
\end{centering}
SN~2015bs was discovered on the 25th of September 2014 by the \textit{Catalina Sky Survey} (CSS) telescope
of the overall CRTS project. In Supplementary Fig.\,1 the position of the SN is indicated on a collapsed
image produced from our integral field spectroscopy obtained using the \textit{Multi Unit Spectroscopic Explorer} (MUSE\cite{bac14}) at the \textit{Very Large Telescope}, VLT.
While no explosion-epoch constraining non-detections 
exist from the CSS, a non-detection from the Mount Lemmon facility (part of CRTS) on the 15th of September
2014 (limiting magnitude of 21.73) constrains the date of explosion to be the 20th of September 2014 $\pm$5 days, or
Julian Date (JD) 2456920.5 $\pm$5 days. 
The transient was also detected much later by Pan-STARRS1 as PS15dsr on the 
27th June 2015, at $w_{\rm ps}$ = 21.3 mag.
A spectrum was obtained with the \textit{ESO Faint Object Spectrograph and Camera (v.2)} (EFOSC2\cite{buz84}) mounted on the \textit{New Technology Telescope} (NTT) 
on September the 29th 2014\cite{wal14} (see Supplementary Fig.\,2), revealing a type II spectral morphology. 
Matching of the classification spectrum with a library of
SN spectral templates using SNID\cite{blo07} gives good results with SN~2004et 
at 2 days before maximum light, which translates to 14 days post explosion (+14\,d), leading to
an explosion epoch of the 15th of September, i.e. the same date as the last non-detection. 
The reason for the earlier explosion epoch from the spectral matching can
be explained by the low progenitor metallicity of SN~2015bs, meaning that spectral line and colour evolution is slower
than usually observed in SNe~II (because of reduced blanketing  by metal lines\cite{des11}).
Nevertheless, the spectral matching gives an explosion epoch that is overall consistent
with that from the non-detection.\\
\indent 
We identify narrow \ha\ emission
from a galaxy (indicated in Supplementary Fig.\,1) offset 3.4\as\ from SN~2015bs (the characterisation of which is presented below). 
The observed wavelength of this emission (the only emission line observed for this galaxy) gives a
redshift for that object of 0.027. This redshift is consistent
with that of the SN \ha\ emission as seen in the nebular spectrum, and the value inferred from spectral matching above. 
We adopt this redshift for SN~2015bs, which corresponds to a distance modulus of 35.4 mag (assuming $H_0$ of 73 km s$^{-1}$ Mpc$^{-1}$).
At this redshift, the angular separation between the peak flux of our identified host and SN~2015bs translates to 2 kpc.\\
\indent Line of sight extinction from dust contained within the Milky Way is taken from the recalibrated dust maps of \cite{sch11},
assuming a Fitzpatrick extinction law\cite{fit99}. No sign of narrow sodium absorption within the spectra
of SN~2015bs is detected -- the presence of which would indicate a higher level of host galaxy extinction -- and
as shown below in Section 3 and Supplementary Fig.\,3, SN~2015bs does not show particularly red colours. Therefore
we assume that SN~2015bs is affected by negligible internal host galaxy extinction.

\vspace{0.5cm}
\begin{centering}
\textbf{2. Data reduction and calibration}\\
\end{centering}
$BVri$ photometry was obtained through the \textit{Carnegie Supernova Project-II} (CSP-II\cite{ham06}) from around maximum light to just after the 
end of the plateau, using the Swope telescope (+ e2V CCD) at the Las Campanas Observatory.
Images were reduced in a standard manner. Observations of standard star fields were carried
out on photometric nights when SN~2015bs was observed allowing the calibration of local standard sequences
in $BV$\cite{lan92} and $ri$\cite{smi02}. Photometry
of SN~2015bs was then calibrated against these local sequences,
and is published in the natural system of the Swope telescope. Photometry of 
local sequence stars is presented in
Supplementary Table\,1 (on the \textit{standard} system), while that of SN~2015bs is listed in Supplementary Table\,2 (on the 
\textit{natural} system). 
No attempts were made to subtract the underlying host galaxy 
flux, given that the host is not detected in deep images taken around a year post explosion. The $BVri$ light-curves
for SN~2015bs are displayed in Supplementary Fig.\,4.\\
\indent The Pan-STARRS Survey for Transients observed the field of SN~2015bs
during  the tail phase, some 280 days after discovery during its normal survey mode.
The transient was recovered over a period of 77 days with the internal name
PS15dsr. The data were taken with the broad $w_{\rm ps}$ band which is a
composite of $g_{\rm ps}r_{\rm ps}i_{\rm ps}$, as described in \cite{cha16}. Difference imaging with respect to a reference frame was carried out,
with point-spread-function photometry produced automatically as described
in \cite{wat16} and \cite{mag16}. The detections from Pan-STARRS
difference images are associated and merged into objects
in a database of transients\cite{sma16} and the photometry is reported in
Supplementary Table\,3 (AB magnitudes in the system described by \cite{ton12})\\
\indent Four photospheric-phase optical spectra were obtained for SN~2015bs using the NTT (+ EFOSC2)
at La Silla, through the PESSTO collaboration, and using the \textit{Las Cumbres Observatory} (LCO\cite{bro13}) FLOYDS spectrograph. 
Spectra were obtained at +9 (the classification spectrum discussed above), 
+23, +57 and +80\,d. The photospheric-phase spectral sequence is presented in Supplementary Fig.\,2. 
EFOSC2 spectra were reduced and calibrated in a standard manner using a custom built pipeline for the 
PESSTO project\cite{sma15b}, while the FLOYDS spectrum was reduced as in \cite{val14}. \\
\indent The position and surrounding sky of SN~2015bs were observed using MUSE at +406 and +421\,d.
MUSE is a 1\am$\times$1\am\ field of view (FOV) integral field spectrograph, allowing us to simultaneously 
observe the SN and search for its host galaxy. These data were reduced using the MUSE pipeline\cite{wei14}, with subsequent combination of the two 
observations.
The extracted 1 dimensional spectrum of SN~2015bs is shown in Fig.~2 of the main article. 
The MUSE data cube was analysed
using QFitsView\cite{ott12}.
\\
\indent Throughout our analysis we compare the properties of SN~2015bs with a sample of SNe~II from the literature.
Given that our conclusions stem from analysis of nebular-epoch optical spectroscopy, 
our comparison sample was defined as any SN~II with a nebular spectrum (with a cut off date of December 2015) 
within $\pm$50 days of that obtained for SN~2015bs, with respect to
the explosion. Seven such SNe were found which are listed in Supplementary Table\,4.

\vspace{0.5cm}
\begin{centering}
\textbf{3. Nebular line analysis of SN~2015bs with respect to a SN~II comparison sample}\\
\end{centering}
Nebular spectra
of SNe~II are dominated by \ha, [\oi] 6300,6364 \AA, and [\caii] 7291,7323 \AA, 
and our analysis is restricted to the measurement of these
line profiles. We measure FWHM velocities,
and in the case of [\oi] and [\caii] their absolute fluxes.
Velocities are extracted by fitting Gaussians
to each line and measuring their FWHM. In the case of [\caii], often more than two Gaussians
are needed to provide a good fit. This is to be expected, as the [\caii]
lines can be blended with e.g. [\niii] 7378 \AA, [\feii] 7388 \AA, and [\niii] 7412 \AA.
In addition, for SN~2015bs and SN~2013ej, more than two components are needed
for [\oi], and more than one component for \ha\ (arguing against unusually strong 
[\niii] 7378 \AA, [\feii] 7388 \AA, and [\niii] 7412 \AA\ in these SNe~II, given that the `red-excess' is not
unique to [\caii]). 
In the case of SN~2013ej, it has been suggested that the nebular lines are best modelled assuming
blue- and red-shifted components of [\oi], \ha, and [\caii]\cite{yua16}.
Additional components on the red side of [\oi] and \ha\ were also observed
for SN~2014G\cite{ter16}, and were argued to be due to circumstellar interaction. 
SN~2015bs displays [\oi], \ha\ and [\caii] emission peaks blue-shifted by 
around 1000\,\kms. Excess flux is also observed as a
red shoulder in emission lines (see Supplementary Figs\,8 and 11). Such profiles could
suggest significant dust extinction in SN~2015bs. Alternatively,
ejecta asymmetries may explain the observed line profiles. Given that we also observe
both blue-shifted emission and a red-shoulder excess for [\caii] 7291,7323 \AA\ suggests that
a strong ejecta asymmetry is most likely, as this line predominantly forms outside
the metal core where any dust would reside.\\
\indent If a better
fit to the line profiles is attained using additional Gaussian components then 
these are added, and the [\oi], \ha, and [\caii] velocities are taken from the largest fitted Gaussian. 
To estimate line fluxes we simply integrate the total emission under the [\oi] and [\caii]
line profiles. This is achieved over a constant wavelength range for all SNe, meaning that
we include any `extra' emission observed in the case of [\oi], and that from
[\niii] 7378 \AA, [\feii] 7388 \AA, and [\niii] 7412 \AA\ in the case of [\caii]. This approach
is preferred, given the uncertain nature of fitting to multiple unresolved lines, and
it also allows for a consistent comparison between all SNe.
In this case the values presented here are not 
immediately comparable to those published elsewhere\cite{mag12}.
\\
\indent Histograms of FWHM velocities of [\oi], \ha, and [\caii] are shown in
Supplementary Fig.\,7. With respect to the comparison sample, SN~2015bs shows nebular-phase  velocities
for [\oi] and [\caii] towards the centre of the distributions. However, SN~2015bs clearly displays the highest 
\ha\ velocity (Supplementary Figs\,8 and 9). Larger nebular-phase velocities are expected in the case of SNe~II with higher helium
core masses\cite{des10}. As discussed in the main article, we make direct comparison of the nebular-phase \ha\ velocity measured for SN~2015bs with those
from the hydrodynamic models of \cite{des10}, constraining the progenitor mass of SN~2015bs to be as high as 25\msun\ (further details below).
The `normal' line velocities for [\oi] 
are to be expected for a larger progenitor mass where a higher percentage of the flux is expected
to arise from the core (i.e. a reduced [\oi] flux from faster moving outer envelope).\\
\indent The nebular-line widths are measured directly from the nebular spectrum. However, in order to extract
a photospheric-phase velocity -- and directly compare observed velocities to model values from \cite{des10} -- 
we use two SNe~II from our comparison sample to aid us. This is because the spectral lines usually used for 
estimating the photospheric velocity, such as \feii\ 5169 \AA\ are weak in the photospheric-phase spectra of SN~2015bs, making their use
unreliable. Therefore, we calibrate the velocity difference between \hb\ and \feii\ 5169 \AA\ for SN~2004et and SN~2013ej (given their similar \hb\ and \ha\ velocities
to SN~2015bs), and apply this difference to the \hb\ velocity for SN~2015bs, obtaining a +50\,d photospheric velocity of 5359\,$\pm$\,392\,\kms.
Using this, together with the nebular-epoch HWHM velocity of \ha\ of 2127\,$\pm$\,308\,\kms, thus constrains (through comparison to hydrodynamic models) the initial progenitor mass of SN~2015bs to be between 20 and 25\msun.\\
\indent In the main article the flux of [\oi] with respect to the flux of that across
the full wavelength range of the MUSE spectrum was presented for SN~2015bs as compared to the same measurement
for our comparison sample and SN~1987A. This showed that SN~2015bs indeed displays stronger [\oi] with respect
to the available energy ($^{56}$Co decay).
Previously, the luminosity of [\oi] as a percentage of the total $^{56}$Co power at the epoch of observations 
has been used as a proxy for progenitor mass through model comparisons\cite{jer15}.
Here we also analyse SN~2015bs in this context. To estimate a $^{56}$Ni mass for SN~2015bs we proceed
with three different methods. Firstly, we extract a synthetic $V$-band magnitude 
from the nebular MUSE spectrum obtained at around +400\,d, which we estimate to 
be 24.11$\pm$0.66 mag.
This magnitude is then converted into a bolometric luminosity, and a $^{56}$Ni mass of 0.048$^{+0.041}_{-0.022}$\msun\ is derived 
assuming full trapping of the radioactive emission by the SN ejecta\cite{ham03}.
Secondly, we integrate the total flux within the MUSE spectrum (4800 to 9300 \AA), together with the `MUSE flux' of a 
spectrum of SN~1987A close in time to that of SN~2015bs\cite{phi90}.
Converting these fluxes to luminosities we then obtain the ratio of SN~2015bs MUSE luminosity to that
of SN~1987A, and using a $^{56}$Ni mass of 0.075 for SN~1987A\cite{arn96}, we arrive at a value
for SN~2015bs of 0.042$^{+0.006}_{-0.007}$\msun.
Finally, a $^{56}$Ni mass is estimated using late-time $w_{\rm ps}$-band
photometry. 
We first fit a straight line to the $w_{\rm ps}$-band photometry, confirming a decline rate consistent with that
expected by the decay of $^{56}$Co. We then extrapolate this (by 50 days) to the epoch at which
there is a spectrum available for SN~1987A. A $^{56}$Ni mass of 0.057$^{+0.003}_{-0.003}$\msun\ for SN~2015bs
is then obtained by scaling the brightness of the SN~2015bs photometry to that of SN~1987A.
Taking an average of these three values we obtain  a $^{56}$Ni mass of 0.049$\pm$0.008\msun.\\
\indent Using the derived $^{56}$Ni mass for SN~2015bs the luminosity of [\oi] is estimated as a percentage of the $^{56}$Co power to be 5.3\%.
This is much higher than any previous SN~II (see Fig.~24 of reference \cite{val16}, where the previous
highest value was less than 4\%), and when compared to model predictions suggests a 
ZAMS mass of $\geq$17-18\msun\ (estimated from figure 24 of \cite{val16}), consistent with the mass estimates 
from [\oi] fluxes as compared to models in the main article using the [\oi] flux compared to the total MUSE flux,
and comparison to such constraints for SN~1987A.\\ 
\indent The overall results from this nebular
analysis are shown in Supplementary Fig.\,10. Here two ratios are plotted vs. each other. 
On the x-axis the ratio of the nebular to photospheric-phase \ha\ velocity is shown. 
This normalises the outer core velocity to the explosion energy\cite{des10}.
On the y-axis we show the ratio of the integrated flux of [\oi] to that of
[\caii].
SN~2015bs falls on the extreme of the distribution of each axis, confirming its uniqueness.
Based on model predictions\cite{des10,jer12}, the simplest explanation is that SN~2015bs was the explosion of 
a massive, 17-25\msun\ progenitor star, 
i.e. \textit{the most massive progenitor star} yet inferred for a SN~II.

\vspace{0.5cm}
\begin{centering}
\textbf{4. Host galaxy identification and characterisation}\\
\end{centering}
There is a faint galaxy 12.7\as\ away from the explosion position of 
SN~2015bs (see Supplementary Fig.\,1),
however this galaxy does not have a published redshift and appears as a candidate SDSS galaxy. 
The initial redshift of 0.021 was taken from the SN spectral matching (see above). 
This implied an absolute $r$-band magnitude of --13.6 mag for the galaxy: already one of the dimmest hosts for a SN~II.
However, in our MUSE observations [\oii] 3727 \AA\ and \hb\ emission are identified for this galaxy 
at a redshift of 0.90, inconsistent with our initial redshift estimate and therefore this galaxy was discarded
as the possible host.
The host of SN~2015bs was identified as a very faint galaxy 
(see Supplementary Fig.\,1) that has narrow \ha\ emission at a wavelength consistent with SN~2015bs. 
Only \ha\ is visible in the spectrum, so
we are unable to constrain the metallicity using emission line diagnostics. This provides a compelling argument for the 
use of SN~II as independent metallicity indicators.
A synthetic $r$-band magnitude was extracted from the host galaxy spectrum and estimated to be
23.3$\pm$0.2 mag. Correcting this for Milky Way extinction, and 
the distance modulus, we obtain an absolute
$r$-band magnitude of --12.2, making the host of SN~2015bs the dimmest host for a 
SN~II in the literature. Aware of the caveats of converting this to a metallicity, 
this implies a chemical abundance of 0.04$\pm$0.05\,\zsun \cite{arc10}, making
SN~2015bs the lowest host metallicity SN~II yet studied. The metallicity error is that of the dispersion 
on the relation between absolute magnitude and metallicity from \cite{arc10}.\\

\begin{centering}
\textbf{Data Availability Statement}\\
\end{centering}
The data that support the plots within this paper and other findings of this study are available from the corresponding author upon reasonable request.
In addition, the PESSTO spectra are available through the PESSTO Spectroscopic Data release 3
(SSDR3), for more information see the PESSTO website (\url{www.pessto.org}), all spectra will also be made available 
on WISeREP: \url{www.weizmann.ac.il/astrophysics/wiserep/}, and photometric measurements are listed in the SI.

\begin{centering}
\textbf{\large{References}}
\end{centering}

\clearpage

\LARGE
\vspace{0.5cm}
\begin{centering}
\textbf{Supplementary Information}\\
\end{centering}
\normalsize

\clearpage
\vspace{0.5cm}
\begin{centering}
\textbf{1. Characterisation of SN~2015bs}\\
\end{centering}
To characterise SN~2015bs 
we use a number of parameters that have been defined to study the $V$-band light-curves of type II supernovae (SNe~II), including
magnitudes at various epochs, decline rates at different epochs, and duration of different
phases\cite{and14}. All measured parameters are listed in Supplementary Table\,5. Also listed in the table are the mean 
values of these same parameters as measured for a large sample of $>$ 100 SNe~II\cite{and14}.
SN~2015bs is a bright 
SN~II, characterised by a relatively flat light-curve, a relatively short plateau duration (Pd), and a relatively high $^{56}$Ni mass. 
(At nebular times, and in the absence of other sources like 
interaction with circumstellar material or radiation from a compact remnant, 
the SN luminosity is powered exclusively by the decay of $^{56}$Co. The decay chain is $^{56}$Ni to $^{56}$Co to $^{56}$Fe, 
with a half-life for $^{56}$Ni and $^{56}$Co of 6.0749\,d and 77.233\,d.). However,
all of the measured parameters fall within the distribution of literature SNe~II. 
This is also seen in the absolute $V$-band light curves plotted in Fig.~1 of the main article.\\
\indent In Supplementary Fig.\,4 we show the $B-V$ and $V-i$ colour curves of SN~2015bs and the comparison sample.
To produce these curves, magnitudes are corrected for both
MW and internal host galaxy extinction (with the latter taken from the references in Supplementary Table\,5). In $B-V$, 
SN~2015bs falls on the blue side of the distribution, however, again it does not seem peculiar
in any way. In $V-i$ SN~2015bs falls within the central part of the distribution.\\
\indent The final comparison we make is through measurements of ejecta expansion velocities. In Supplementary Fig.\,5
we plot the time evolution of \ha\ (from the Full Width Half Maximum, FWHM, of the line profile) and \hb\ (from the minimum of the absorption trough) spectral velocities
for SN~2015bs together with those for the SN~II comparison sample. SN~2015bs displays some of the highest velocities,
however their values and time evolution fall within the observed range of SNe~II. One may speculate that 
the lack of strong metal lines may be an effect of high expansion velocities blending relatively weak metal lines into the continuum.
However, even in high energy photospheric models such lines are clearly visible\cite{des13}, and such a scenario does not seem to be at play 
in the case of SN~2015bs.\\
\indent After the characterisation and comparison provided above, we conclude that SN~2015bs is a relatively 
normal SN~II. This event does not appear to be peculiar during the photospheric phase (except for the strength 
of metal lines, as outlined below).
Following this, the progenitor of SN~2015bs was most likely a red supergiant (RSG) star,
consistent with literature constraints on other SNe~II.\\

\vspace{0.5cm}
\begin{centering}
\textbf{1.1 The lack of metal lines in the +57\,d spectrum}\\
\end{centering}
The photospheric-phase spectra of SN~2015bs, as presented in Supplementary Fig.\,2, show a clear lack of metal lines.
This is particularly apparent blue ward of \ha\, and in the `cleanness' of the full Balmer series. 
To our knowledge, such a spectrum has not been observed previously.
In Supplementary Fig.\,6 we present a comparison of the +57\,d SN~2015bs spectrum with those from \cite{tad16},
with the latter SNe~II showing the lowest \feii\ 5018 \AA\ pEWs within their sample. Compared to the SNe~II from \cite{tad16},
SN~2015bs displays a remarkable similarity to the 0.1\zsun\ model spectrum. The comparison SNe~II in Supplementary Fig.\,6 
indeed show relatively weak lines, but none display such a similarity to the 0.1\zsun\ model as SN~2015bs, and additionally
these comparison SNe~II display properties marking them out as abnormal events, while SN~2015bs appears as a standard SN~II 
with its metal lines
absent.\\
\indent Metal line strength in photospheric-phase spectra of SNe~II was first predicted\cite{des13}, and then
shown observationally\cite{and16}, to be dependent on progenitor metallicity. The appearance of the SN~2015bs +57\,d spectrum
suggests a low-metallicity progenitor. Fig.~3 in the main article presented a 
comparison between the SN~2015bs +57\,d spectrum and 0.1\zsun\ and 0.4\zsun\ model spectra, 
together with an example of a probable higher progenitor metallicity observed SN~II in comparison 
to a 2\zsun\ model. The match between SN~2015bs and the 0.1\zsun\ model
is remarkable, given that the models were not tailored
to fit any SN in particular. Measuring the \feii\ 5018 \AA\ pEW in the +57\,d spectrum we obtain 4.25$\pm$0.54 \AA. 
Comparing this to a large sample of such measurements\cite{and14}, SN~2015bs has the 2nd lowest 
pEW with respect to all SNe~II at +50\,d. The lowest, SN~2005dn has an pEW of 4.1$\pm$0.7 \AA. 
The estimated oxygen abundance for the host \hii\ region of SN~2005dn
is $<$8.15 dex (on the N2 scale\cite{mar13}). SN~2005dn is a somewhat fast decliner, with an $s_2$ value of 
1.55 mag 100 days$^{-1}$. Another key difference between SN~2005dn and SN~2015bs is that while SN~2005dn has a low
\feii\ 5018 \AA\ pEW, it displays many additional stronger lines between \ha\ and \feii\ 5018 \AA\ pEW
that are almost non-existent in SN~2015bs (see Supplementary Fig.\,2).
There are two other SNe with pEWs less than 7\AA. These are SN~2006Y and SN~2008gr. 
Both of these SNe are fast decliners and would not be considered 
`plateau' SNe~II. SN~2008gr, in a similar fashion to SN~2005dn discussed above, presents 
numerous other metal lines blue ward of \ha. In the case of SN~2006Y, this is a very non-standard
SN~II and hence any comparison to our `normal' SN~II 2015bs is not particularly useful. The spectra
of low-\feii\ 5018 \AA\ pEW SNe~II presented by \cite{tad16} also display different properties as compared 
to the 0.1\,\zsun\ model and SN~2015bs, as shown in Supplementary Fig.\,6 (with the latter two being 
remarkably similar).\\
\indent In conclusion, SN~2015bs displays the closest resemblance of any SN~II in the
literature to the low-metallicity 0.1\,\zsun\ model, presenting an incredibly `clean' spectrum between \ha\ and \feii\ 5018 \AA.
Based solely on the photospheric-phase spectrum of SN~2015bs, this is the clearest candidate yet for a $\leq$0.1\zsun\ 
progenitor SN~II. (The SN~II, LSQ13fn, was claimed to arise from a $\sim$ 0.1\zsun\
progenitor\cite{pol16}. However, LSQ13fn was a non-standard SN~II -- it was particularly blue, it does not follow the 
magnitude--velocity relation -- while still displaying significant differences with respect to the 0.1\zsun\ model.) 
This conclusion is confirmed by our host galaxy analysis, and further supports
the use of SN~II spectra as metallicity indicators, especially in the case when no host galaxy constraints are available.

\vspace{0.5cm}
\begin{centering}
\textbf{2. Progenitor mass constraints from nebular-line analysis}\\
\end{centering}
Most of our conclusions are grounded on models that predict a strong relationship between
helium core mass (the property being probed by nebular-phase spectroscopy), and Zero Age Main Sequence (ZAMS) mass\cite{woo95}.
However, stellar rotation is known to affect massive star evolution\cite{mey00}. 
In particular, models with higher initial rotation rates tend to produce more massive helium cores\cite{hir04}. While observationally such rotational effects
are unconstrained for SNe~II, in the case of SN~2015bs the progenitor may be affected by different degrees of rotation, 
hence affecting the properties of the resulting SN. There is some 
evidence that lower-metallicity massive stars have higher rotational velocities\cite{hun08} (but with large
dispersion and many low-rotation stars being also found within low-metallicity environments). It is therefore a possibility 
that the high core mass estimated for SN~2015bs is a product of a lower ZAMS progenitor than usual, due to
a higher initial rotation. However, while we cannot rule out this possibility, currently this is somewhat speculative.
An additional uncertainty lies with the treatment of convection in 1-D (spherically symmetric) stellar evolution
models\cite{arn11}, which
may lead to uncertainty in the exact mapping of helium-core mass back to ZAMS.
Stellar evolution models that include rotation and/or a 3-D treatment of convection etc are lacking, and, 
therefore, spectral models based on such progenitors are also lacking.
For nebular 
models that do exist\cite{des10,jer14},
SN~2015bs is consistent with a relatively massive progenitor.
In massive stars the helium core is formed during the hydrogen-burning phase\cite{woo02}, which
is largely over when the star becomes a RSG. The bulk of the mass loss by a 15-30\msun\ star, during the RSG phase 
and thus after the helium core is formed\cite{woo02}. 
Therefore uncertainties in RSG mass-loss rates (and their dependence on metallicity and rotation) have little impact 
on our helium core mass constraints for SN~2015bs. 
The effect of progenitor metallicity on helium core mass is indeed found to be negligible up to
an initial mass of about 28\msun\cite{woo02}, and this is also illustrated by simulations\cite{des13}. 
15\msun\ stars of 0.002, 0.008, 0.02 and 0.008 metallicity evolved using standard stellar evolution prescriptions
(in particular the dependence of mass loss on metallicity) and exploded to produce transients with light-curves and
spectra consistent with standard SNe~II produced helium core masses that range from 4.15\msun\ for the lowest 
to 3.77\msun\ for the highest metallicity cases\cite{des13}. Therefore, only a 10\%\ change in helium core mass is observed from a
change of a factor of 20 in metallicity. This change is too small to explain the unique properties of the nebular-phase properties 
of SN~2015bs. 
While future nebular-phase spectral modelling using a large range of progenitor properties will certainly be useful for 
this type of study, 
comparison with current models suggest a relatively large ZAMS for SN~2015bs.

\vspace{0.5cm}
\begin{centering}
\textbf{3. Comparison to Jerkstrand models and SN~1987A}\\
\end{centering}
Above we analysed and discussed specific line measurements of SN~2015bs in its nebular-phase spectrum and
their implications for the helium core-mass and progenitor mass. In this sub-section we directly compare our nebular spectrum
to a) model nebular spectra, and b) a nebular spectrum of the intensely studied SN~II 1987A.\\
\indent Supplementary Fig.\,11 shows a comparison between the 15 and 25\msun\ ZAMS Jerkstrand models\cite{jer14}
at +400\,d, with our nebular spectrum of SN~2015bs.
The observed and model spectra display reasonable agreement overall. 
The strength of the [\caii] 7291,7323 \AA\ lines are very similar, while the \ha\ flux is considerably higher in the
15\msun\ model as compared to SN~2015bs.
SN~2015bs falls between the 15 and 25\msun\ models with respect to the relative strength of [\oi]. 
While one may therefore conclude that this implies a $\sim$20\msun\ ZAMS mass for SN~2015bs, within the models
the strength of [\oi] does not increase linearly between 15 and 25\msun \cite{jer14} (unfortunately
there is no 19\msun\ model at +400\,d). Compared to this set of models, SN~2015bs is therefore
consistent with a progenitor ZAMS mass of 17-18\msun, although as above we note that model
line fluxes start to saturate above 19\msun, meaning
that models in the 20-30\msun\ range are only 20-30 brighter than the measured luminosity, and
cannot be ruled out considering uncertainties in the modelling.\\
\indent SN~1987A remains the closest observed SN to Earth to explode within the last $\sim$400
years. While the light-curve displayed a rare long rise to maximum, and therefore the event itself
is not directly comparable to the `normal' SNe~II discussed here, given the wealth of
information we have on this event and its progenitor, a direct comparison between SN~1987A and SN~2015bs is warranted.
Many progenitor models have been produced for SN~1987A\cite{utr15}, however we still lack an adequate model
that explains all properties of the observed progenitor, the SN light-curve and its spectral morphology\cite{utr15}.
Models in best agreement with observations of SN~1987A and its progenitor fall in the range 15 to 
20\msun \cite{utr15,suk16,sma09b}. With these constraints in mind a direct comparison
of our nebular spectrum of SN~2015bs with that of SN~1987A at a similar epoch is presented in Supplementary Fig.\,12.
Overall the spectra appear quite similar, however the relative strength of the nebular lines is somewhat different.
SN~2015bs displays stronger [\oi] but weaker [\caii] and \ha\ than SN~1987A. Taking the strength of the [\oi]
as a direct He-core and therefore ZAMS progenitor mass indicator, this suggests that SN~2015bs arose from a higher
mass progenitor than SN~1987A. Given the relatively large progenitor masses (as compared to the current 
population of direct progenitor detections for `normal' SNe~II) discussed for SN~1987A, this comparison 
brings further support to our conclusion that SN~2015bs arose from the highest progenitor yet observed 
for a SN~II, with an overall range of 17-25\msun. Indeed, the direct comparison here with
SN~1987A would suggest masses in the higher end of this range.

\vspace{0.5cm}
\begin{centering}
\textbf{4. The red supergiant problem}\\
\end{centering}
Our interest in the discovery of a high-mass progenitor for a SN~II is motivated by
the previous apparent non-detection of high mass progenitors associated with SN~II explosions.
This `red supergiant problem' refers to the lack of SN~II progenitors with masses higher than $\sim$18\msun\cite{sma15}.
This is a problem because RSGs of up to 30\msun\ are observed\cite{lev05} and at least some are
expected to explode as SNe~II, as they probably lose insufficient mass 
during their late phases to explode as SNe of types IIb, Ib or Ic (`stripped-envelope SNe',
SE-SNe).
However, model predictions for the mass-limit for producing SNe~II or SE-SNe are heavily dependent 
on the mass-loss rates employed and the inclusion and degree of rotation assumed\cite{hir04,mau11,chi13,ren17}, and
it has been claimed that the RSG problem simply does not exist\cite{chi13} when using sufficiently high mass-loss rates\cite{van05}.
If one assumes significant rotation and/or high mass-loss rates, then 
stars initially more massive than $\sim$18\msun\ may evolve to explode as non-SN~II types.
RSG mass-loss rates are hugely uncertain (see e.g. \cite{smi14} for a review),
with significant differences between different recipes\cite{van05,mau11}. High mass-loss rate prescriptions\cite{van05},
have been used in stellar evolution models to demonstrate that $<$20\msun\ stars can explode
as e.g. SNe~IIb\cite{geo12}. However, a consensus to apply such large mass-loss rates to all RSGs and to the overall evolution
of these stars (in place of specific phases of possibly enhanced mass-loss) is still lacking, 
and there is no general agreement on which mass-loss rates one should employ.
While it is possible to produce SE-SNe for $\sim$20\msun\ progenitors when using the highest RSG mass-loss
rates, there are many RSGs that are observed to lose mass at a much lower rate and hence may
not lose enough mass to avoid exploding as a type II SN (see e.g. figure 1 in \cite{mey15} for a comparison
of different mass-loss estimates for RSGs).\\
\indent This red supergiant problem also emerges from nebular spectroscopic studies\cite{jer14}, together with
the analysis of parent stellar populations of SN remnants\cite{jen14}. 
We have also not detected \textit{any} clear cases of 
SNe types II, IIb, or Ib with
progenitors of 20-30\msun. If $>$18\msun\ progenitors explode as non-SNe~II then we should have detected their progenitors.
Both constraints on SNe~IIb and Ib ejecta masses\cite{dro11,can13,lym16,tad18}, and their relative rates\cite{smi11} 
suggest that the majority of these explosions
do not arise from $>$20\msun\ explosions, but rather lower-mass progenitors that lose their mass through binary interaction. 
It has been argued\cite{wal12} that progenitors could be affected by unaccounted for circumstellar dust, therefore lowering 
their progenitor luminosities and estimated masses. However, in the particular case of SN~2012aw the
correct treatment of the dust revealed a relatively low progenitor mass\cite{koc12} (after initially higher estimates). 
If unaccounted for circumstellar dust is affecting derived progenitor luminosities, then an interesting possibility
is that this preferentially affects the higher-metallicity cases where it is assumed that dust production is easier. This could
then be consistent with our conclusion of a low-metallicity--high-mass progenitor for SN~2015bs.\\
\indent An overall relatively low-mass progenitor population for SNe~II is also consistent with the low x-ray luminosities
measured for SNe~II\cite{dwa14} (see \cite{sma15} for detailed discussion of all these possibilities and why -- in their view -- they are unlikely).
Most recently, it has been claimed\cite{dav18} that the lack of high-mass RSGs exploding as SNe~II is due to the incorrect
use of fixed bolometric corrections for estimating progenitor luminosities. \cite{dav18} argue
that bolometric corrections for RSGs change as stars come closer to core-collapse and that when this is taken into account
the observed upper limit of progenitors comes into line with observed RSG masses in the local Universe.\\
\indent The above discussion highlights the current problems in trying to understand the RSG problem, with explanations
ranging from: `there is no problem, $>$20\msun\ stars don't explode as RSGs at solar metallicity because of mass-loss', or
`there is no problem as bolometric corrections are incorrectly applied to obtain progenitor luminosities' to the original
claim that there are RSGs that \textit{should} explode as SNe~II but don't, which together with the lack of detected SE-SNe arising from
$>$20\msun\ progenitors leads to a real problem of a lack of massive star explosions.\\
\indent One possibility -- whether the RSG problem exists or not -- is that the majority of massive progenitors do not explode as 
successful SNe, but instead collapse
to black holes with weak or no accompanying outburst\cite{suk16}. Several independent studies have discussed the `compactness' parameter (basically the core radius within which
a certain mass is confined) as determining whether a successful explosion will occur\cite{oco11,ugl12,suk14}.
These studies suggest that it is indeed more difficult to explode more massive stars, with only small
regions of the $>$20\msun\ progenitor parameter space producing successful explosions, i.e. the `islands of explodability'
(see \cite{mue12} for a successful explosions of a 27\msun\ star). 
The case for failed explosions above 18-20\msun\ is also supported by the compact remnant mass function that is found to 
be bimodal (between neutron stars and black holes). A continuous function would be expected for a continuous range 
of successful explosions\cite{fry12}, however the observed bimodal mass function is a direct prediction of failed 
explosions\cite{koc14}. Finally, the discovery of the disappearance of a $\sim$25\msun\ RSG has provided direct evidence 
for a failed SN\cite{ada17} (see also \cite{rey15}). This gives one specific example of a massive RSG that died with a massive and extended 
hydrogen-rich envelope but did not explode as a luminous transient.\\ 
\indent It is in the above context that our conclusions
of a massive 17-25\msun\ ZAMS mass progenitor for SN~2015bs are placed. Our detection of a massive RSG exploding 
as a SN~II, together with the failed explosion of a similarly massive star\cite{ada17}, suggest that
predicted `islands of explodability' may indeed exist. At the same time, the low-metallicity nature
of SN~2015bs possibly suggests that a prerequisite for
SN~II explosions from progenitors above 18-20\msun\ may be low metallicity, with higher metallicity progenitors
exploding as SE-SNe. However, the detection of a disappearing $\sim$25\msun\ RSG at solar metallicity suggests that
this cannot be the case for \textit{all} RSGs. A high-mass SN~IIb or SN~Ib progenitor at solar-metallicity is required to support this latter scenario.

\begin{centering}
\textbf{\large{References}}
\end{centering}

\begin{suppfigure}
\includegraphics[width=16cm]{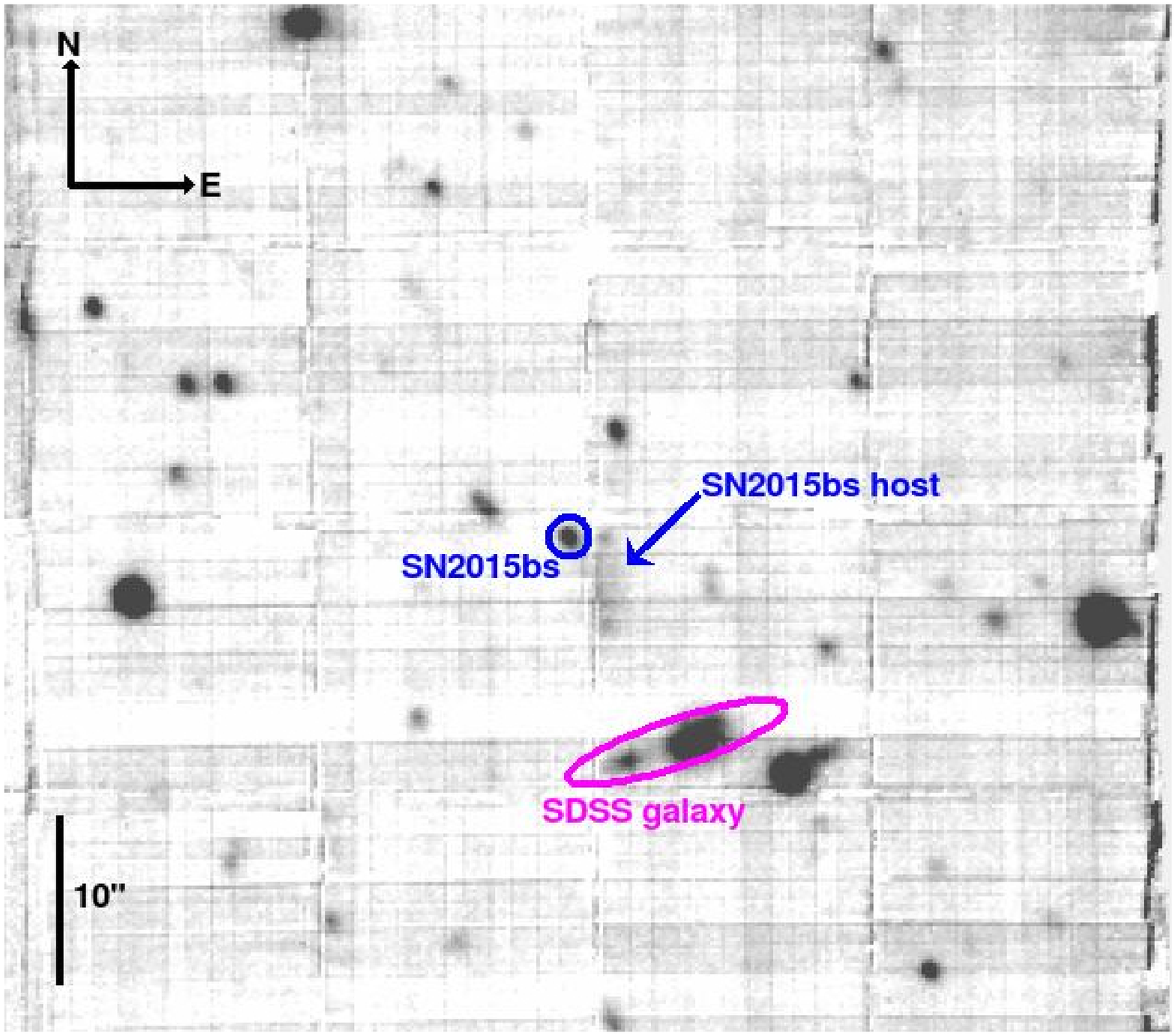}
\caption{MUSE image of SN~2015bs and its surroundings. This is formed 
by collapsing the data cube across the full wavelength range of the observation.
SN~2015bs together with its host and the originally identified possible (SDSS) host are indicated on the image.
The hatched line pattern observed is due to the edges of the individual IFUs within MUSE.}
\label{finderMUSE}
\end{suppfigure}

\begin{suppfigure}
%\begin{centering}
\includegraphics[width=14cm]{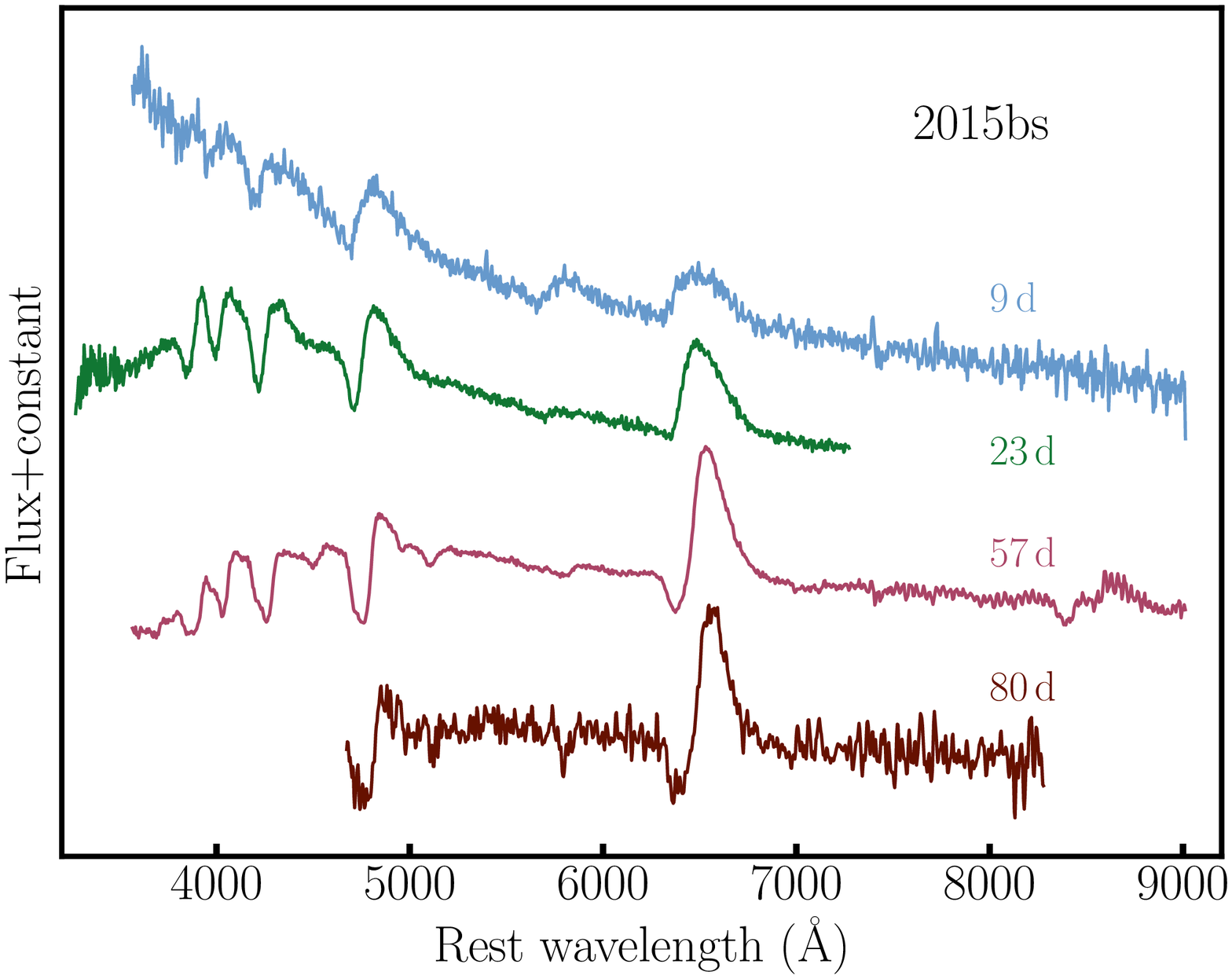}
\caption{Photospheric phase spectral observations of SN~2015bs. The epochs post explosion are
indicated on the plot.}
%\end{centering}
\label{specseq}
\end{suppfigure}

\begin{suppfigure}
\includegraphics[width=12cm]{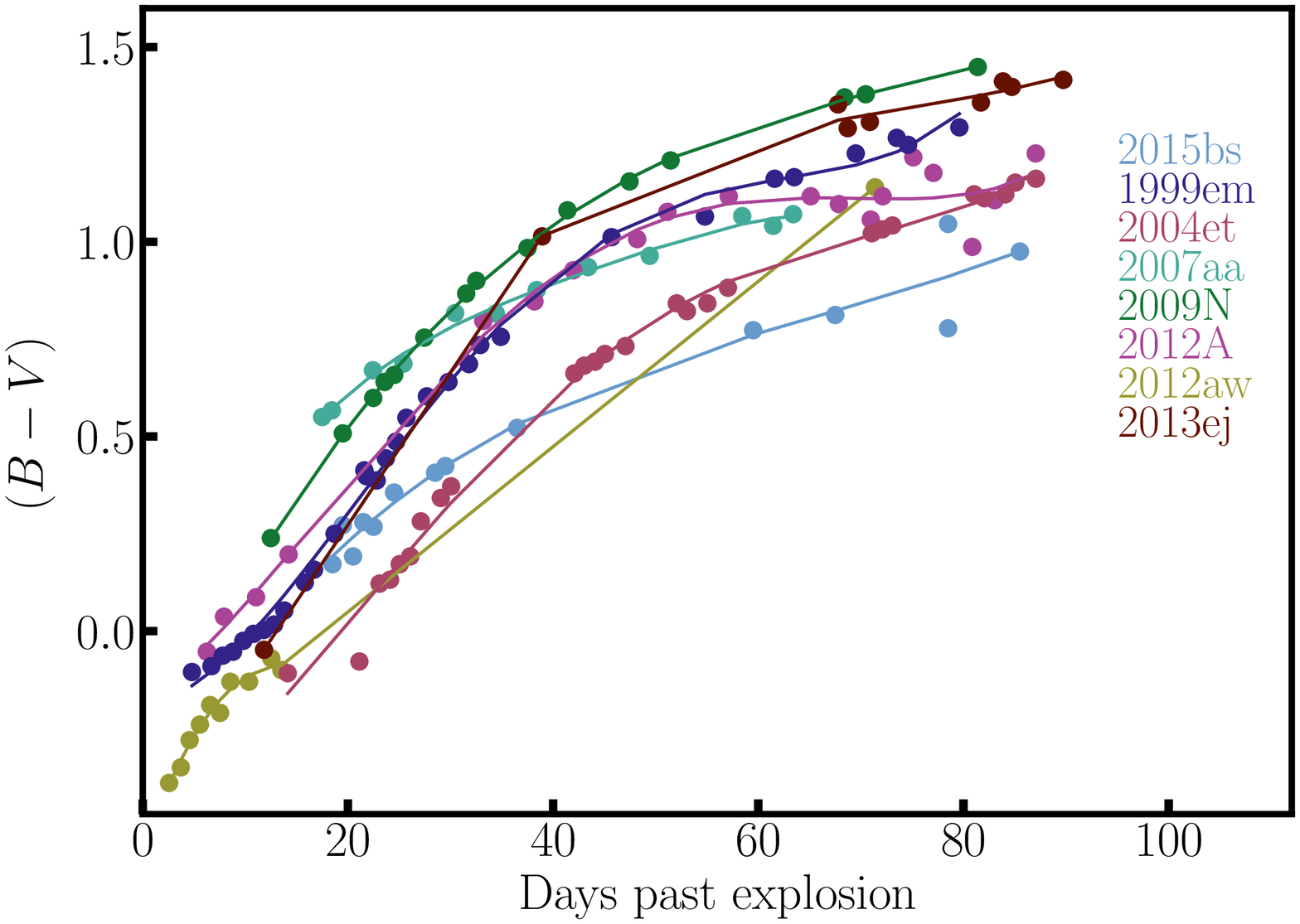}
\includegraphics[width=12cm]{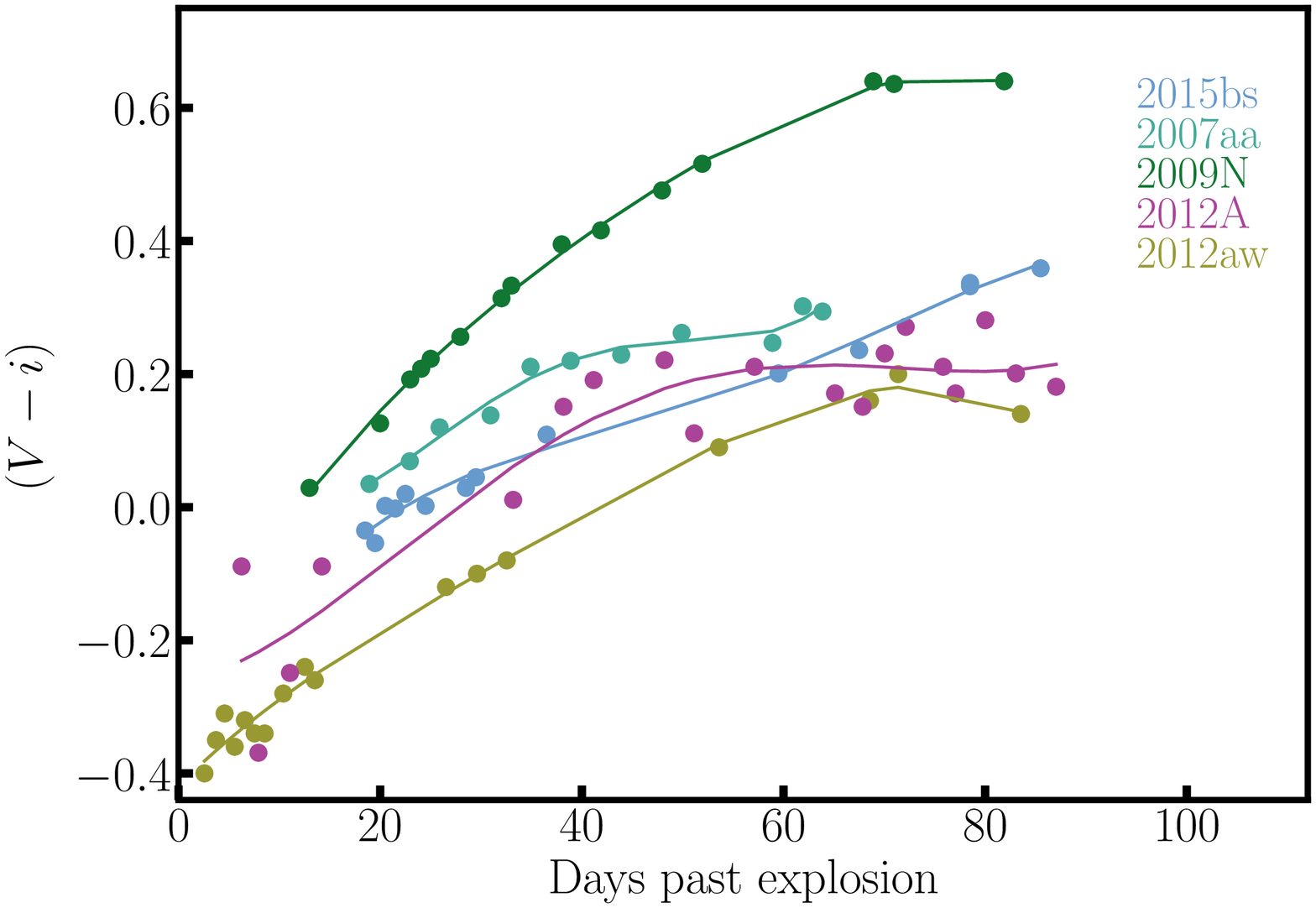}
\caption{$B-V$ (top) and $V-i$ (bottom) colour curves for SN~2015bs and our comparison sample.}
\label{colcurve}
\end{suppfigure}

\begin{suppfigure}
\includegraphics[width=14cm]{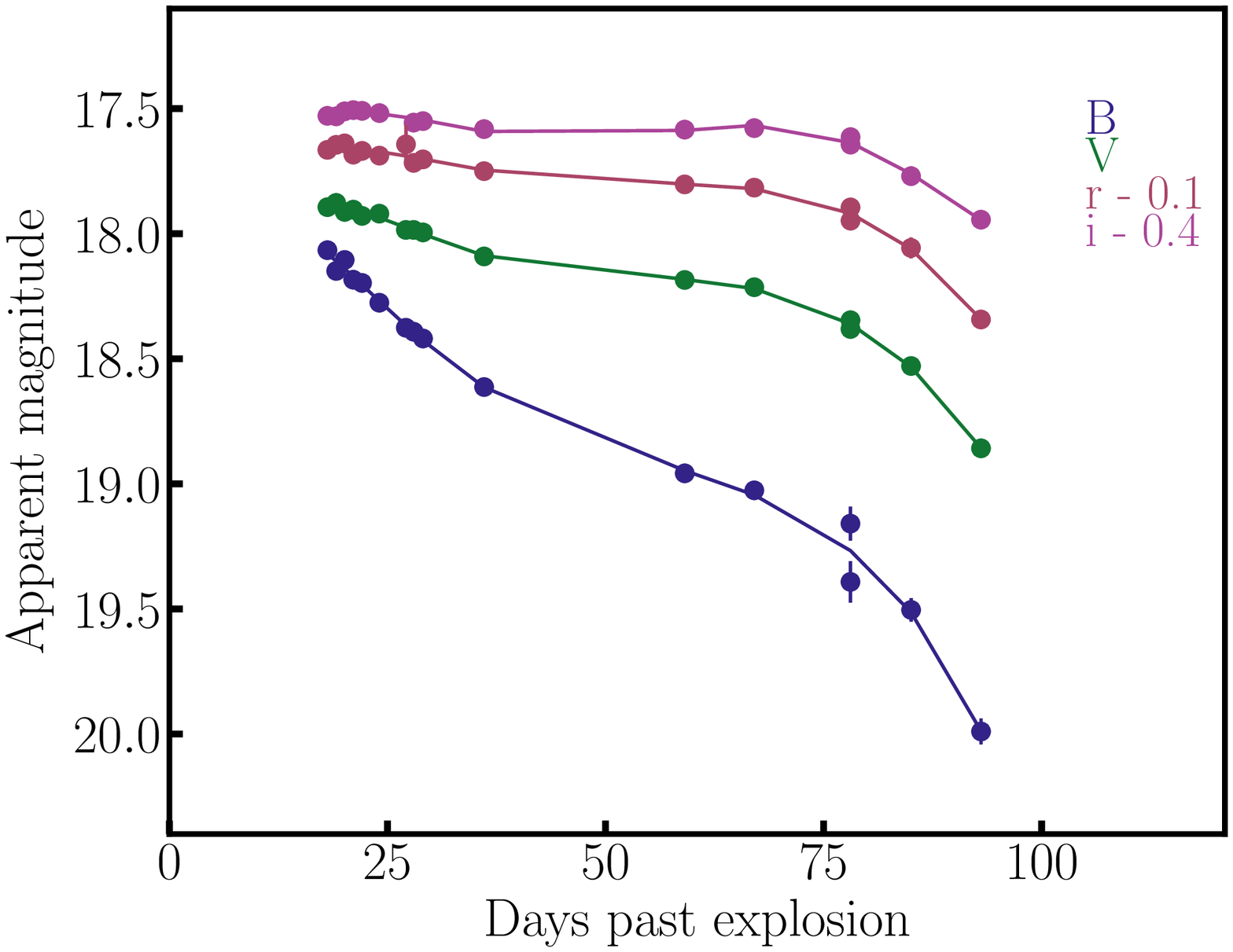}
\caption{$BVri$ light-curves of SN~2015bs. Photometric errors are the propagated errors from the photometric calibration.}
\label{lc}
\end{suppfigure}

\begin{suppfigure}
\includegraphics[width=14cm]{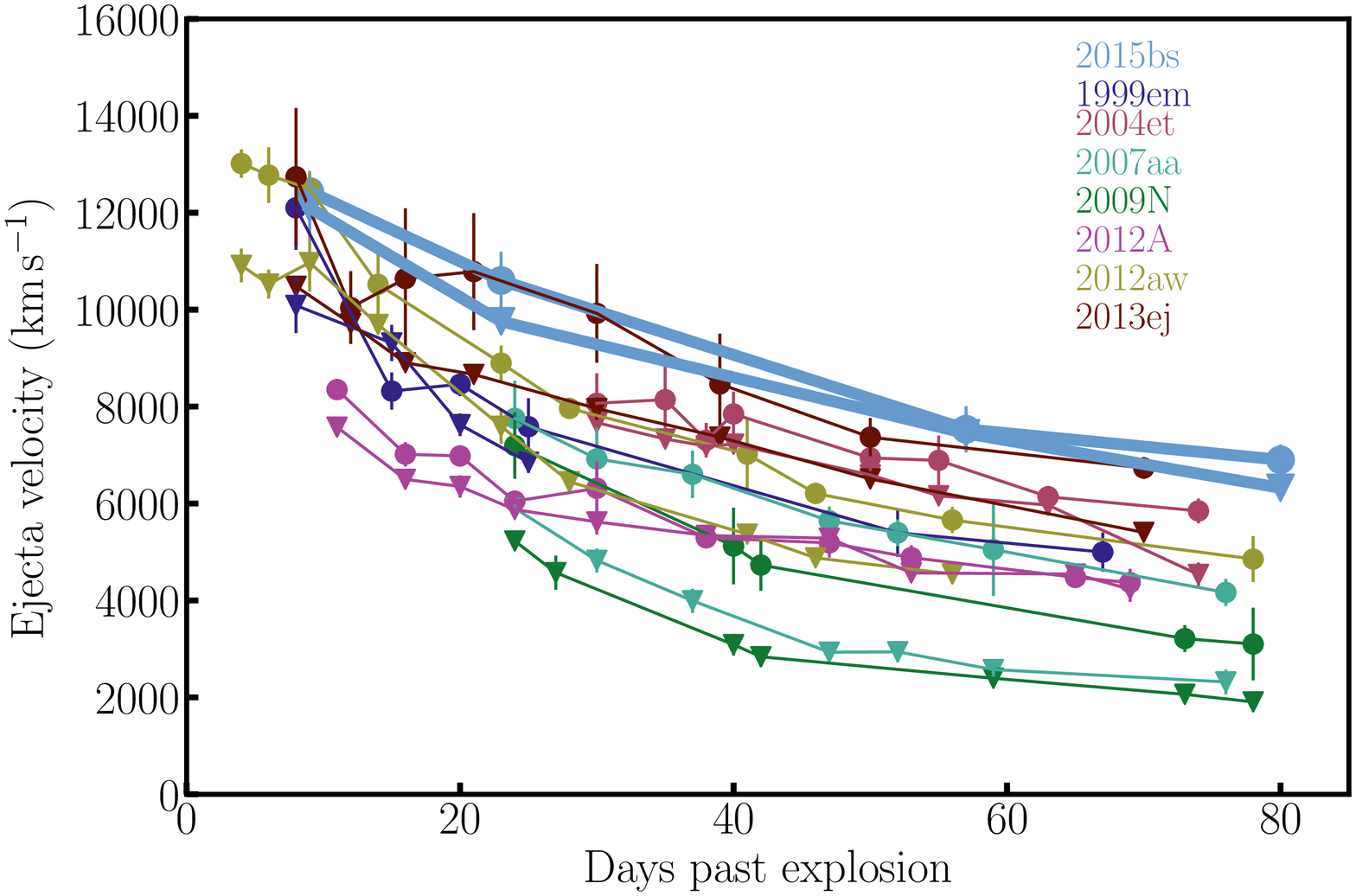}
\caption{\ha\ FWHM (shown as circles), and \hb\ absorption minimum (presented as triangles) 
ejecta velocities for SN~2015bs and the SN~II comparison sample.
Velocity errors are the standard deviation of multiple spectral measurements with slight changes in the 
defined continuum level.}
\label{specvelevol}
\end{suppfigure}

\begin{suppfigure}
\includegraphics[width=15.9cm]{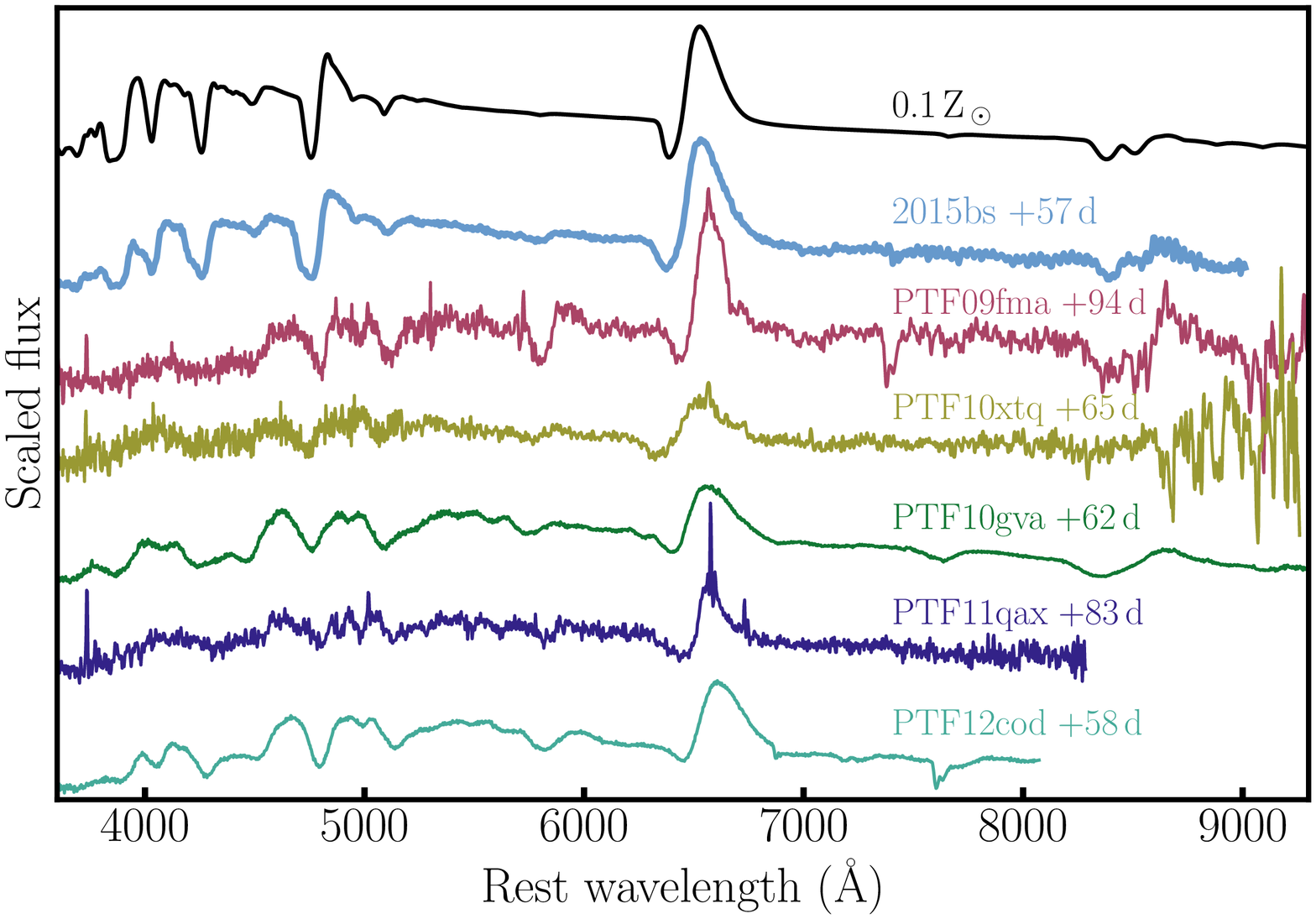}
\caption{Comparison of the +57\,d spectrum of SN~2015bs with that of the 0.1\,\zsun\ model, together
with a number of other SNe~II displaying weak \feii\ 5018 \AA\ lines in a region where line blanketing is strong
at the recombination epoch in \zsun\ SNe~II models and observations.}
\label{iptf}
\end{suppfigure}

\begin{suppfigure}
\includegraphics[width=6.5cm]{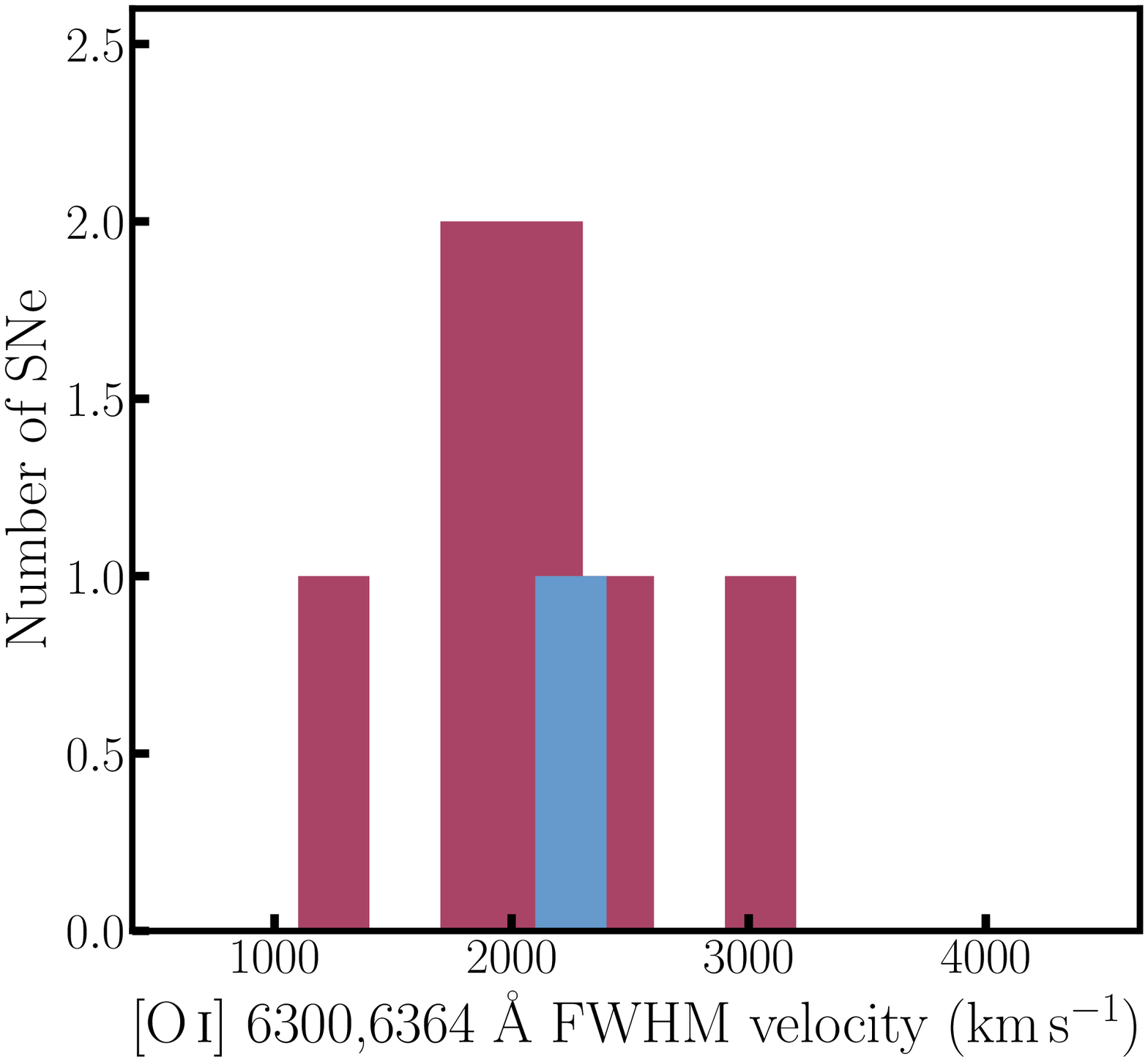}\\
\includegraphics[width=6.5cm]{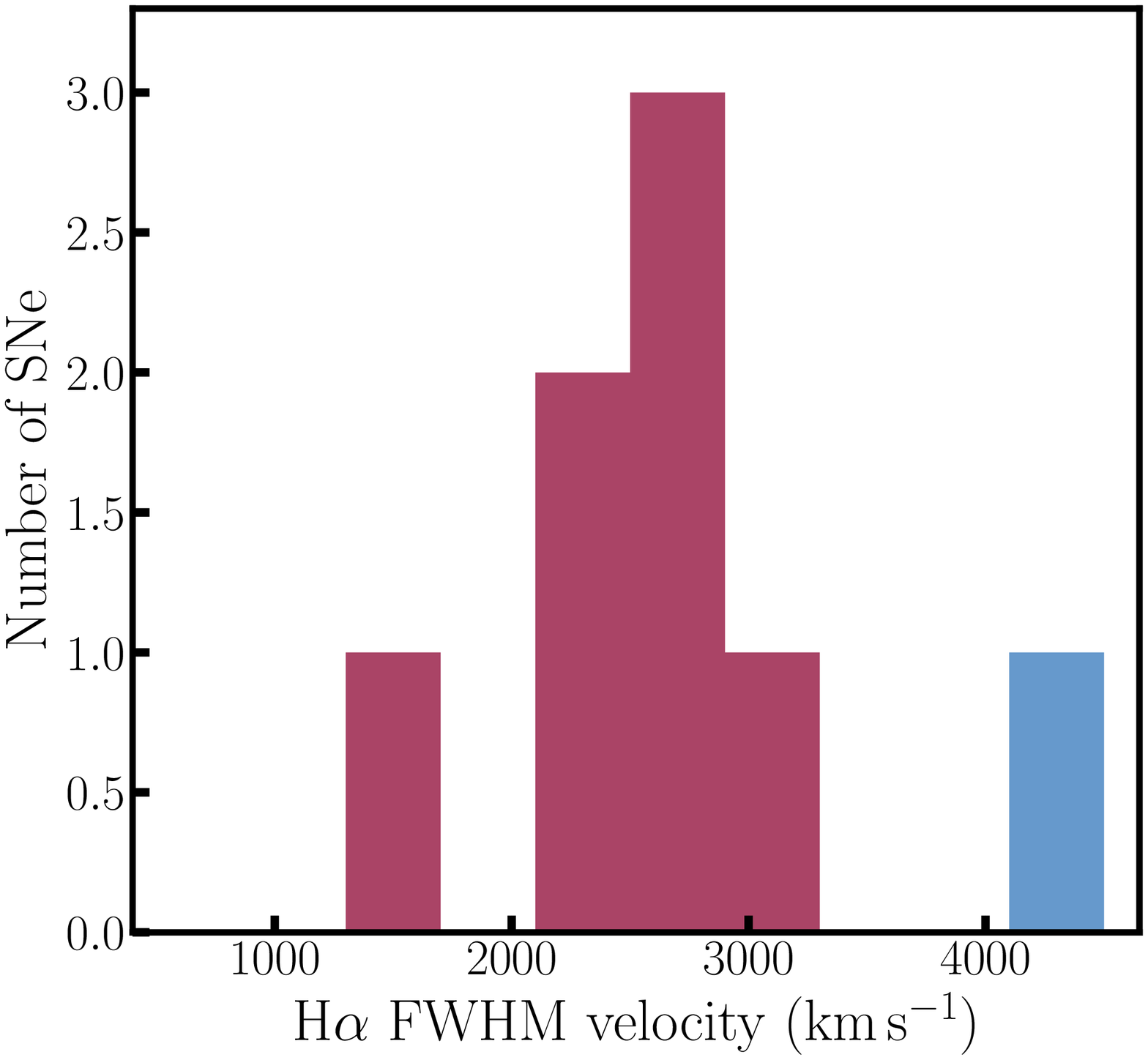}\\
\includegraphics[width=6.5cm]{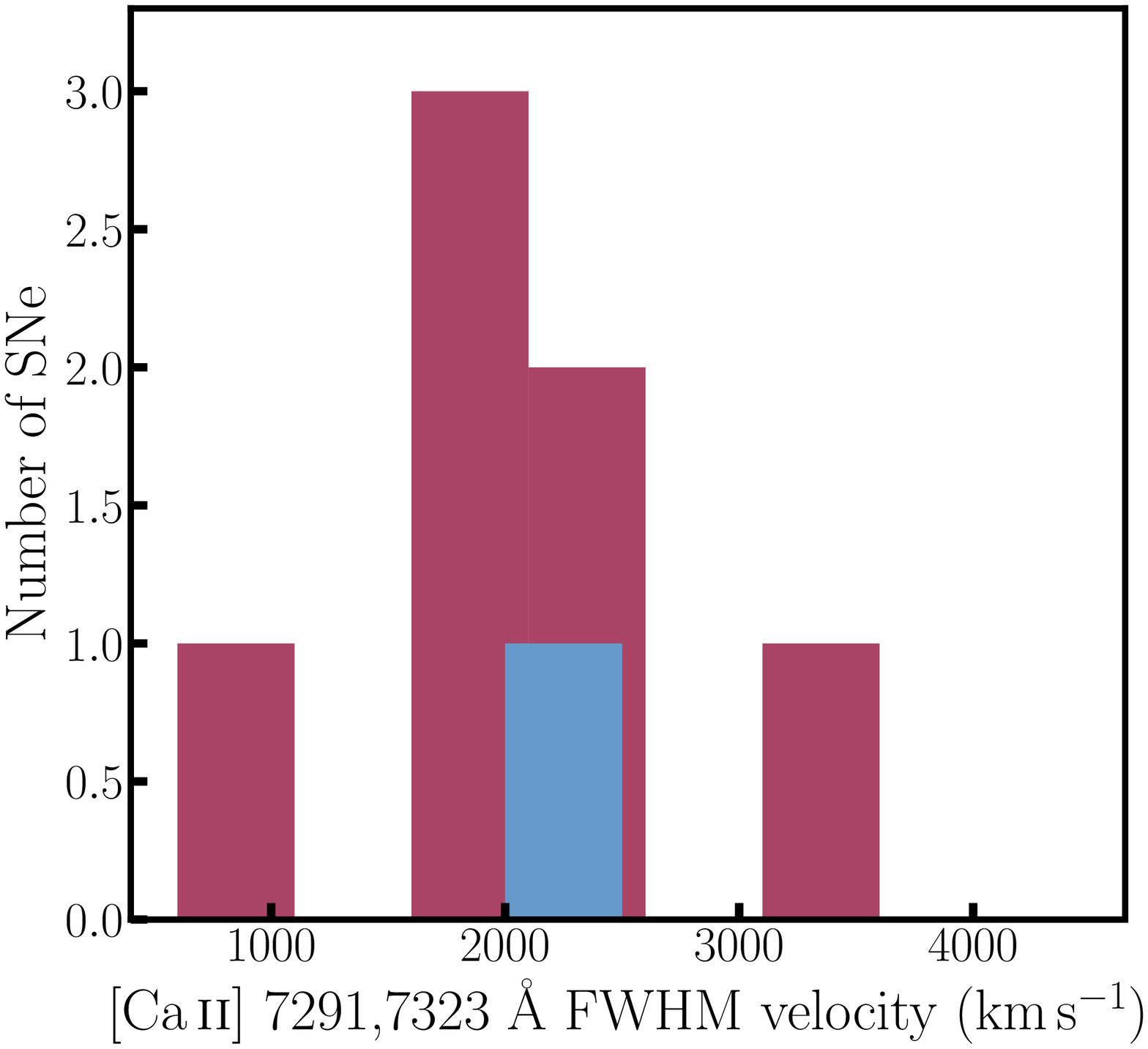}\\
\caption{Histograms of the FWHM velocities of [\oi] 6300,6364 \AA, \ha\ and [\caii] 7291,7323 \AA. In each plot the position 
of SN~2015bs is indicated in blue, while the nebular SN~II comparison sample is presented in maroon.}
\label{nebvels}
\end{suppfigure}

\begin{suppfigure}
\includegraphics[width=16cm]{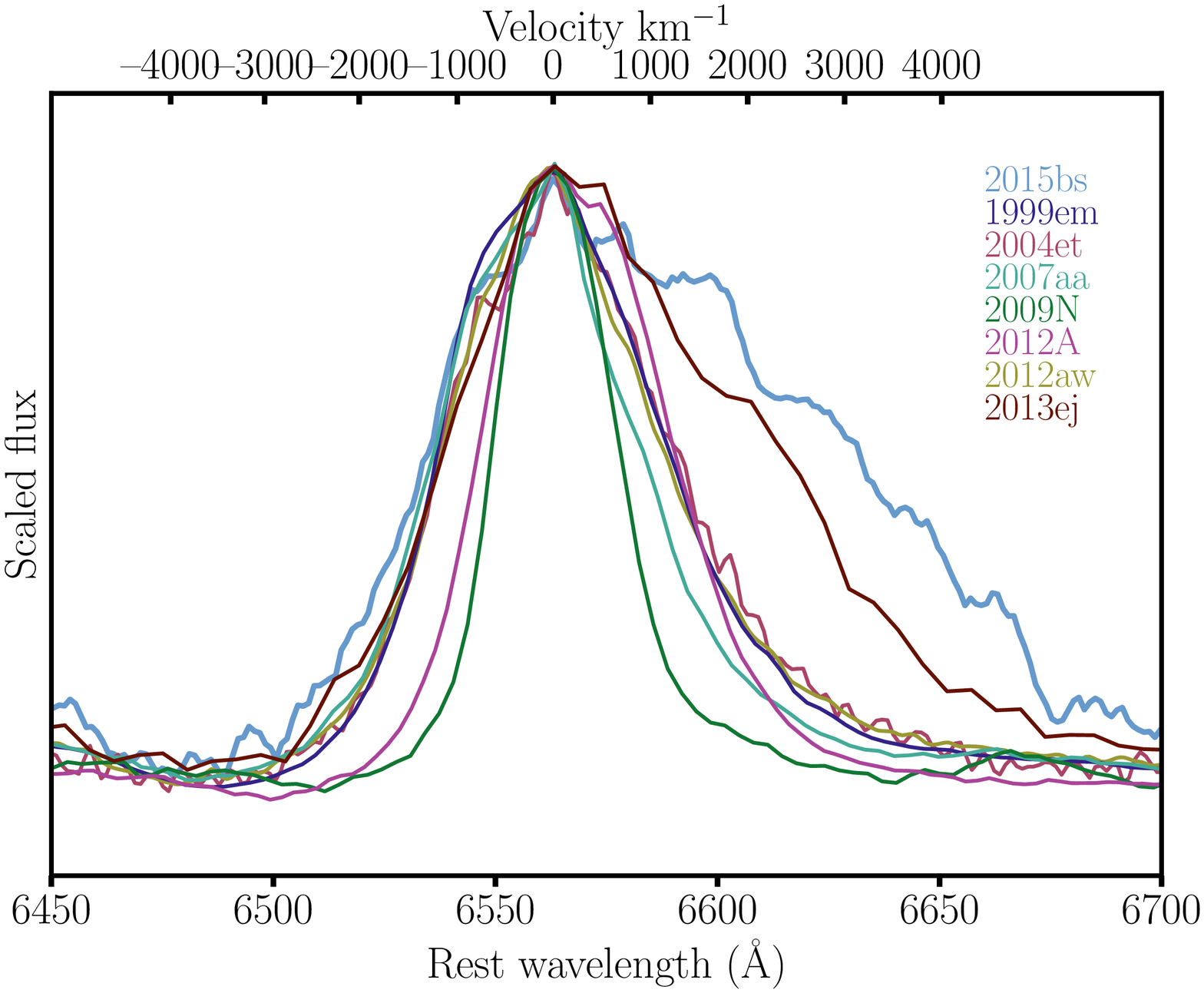}
\caption{Nebular phase \ha\ profile for SN~2015bs and all comparison SNe. Line profiles
have been shifted in wavelength so that the peak of \ha\ aligns for all SNe, and are normalised
to the peak \ha\ flux of SN~2015bs.}
\label{nebHa}
\end{suppfigure}

\begin{suppfigure}
\includegraphics[width=16cm]{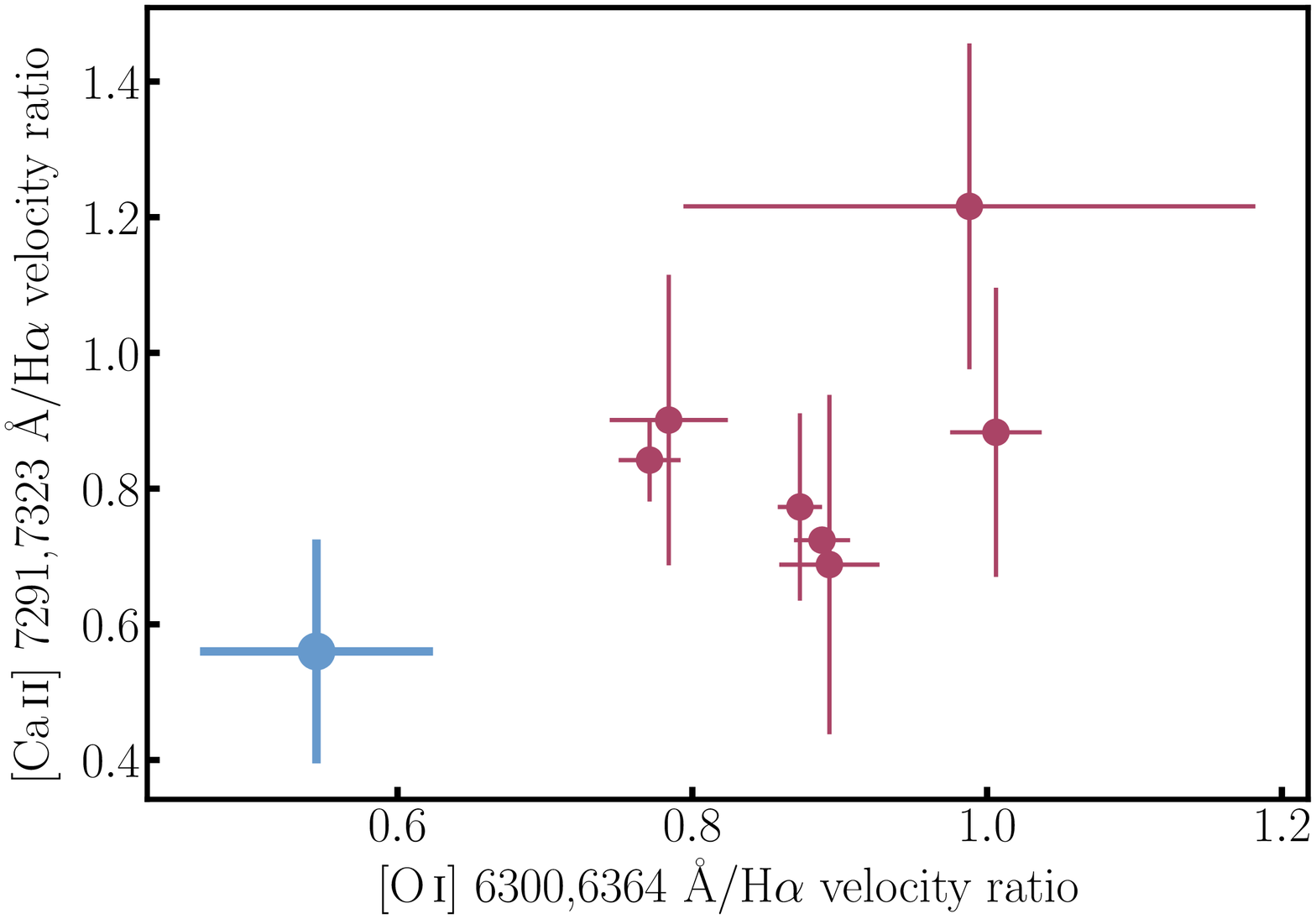}
\caption{Ratio of the FWHM velocities of [\oi] 6300,6364 \AA\ to \ha\ on the x-axis, compared with [\caii] 7291,7323 \AA\ to \ha\ on 
the y-axis. SN~2015bs is presented in blue, with the comparison sample in maroon.
Ratio errors are propagated errors from the standard deviation of multiple spectral measurements of each feature with slight changes in the 
defined continuum level each time.}
\label{nebHavelratio}
\end{suppfigure}

\begin{suppfigure}
\includegraphics[width=16cm]{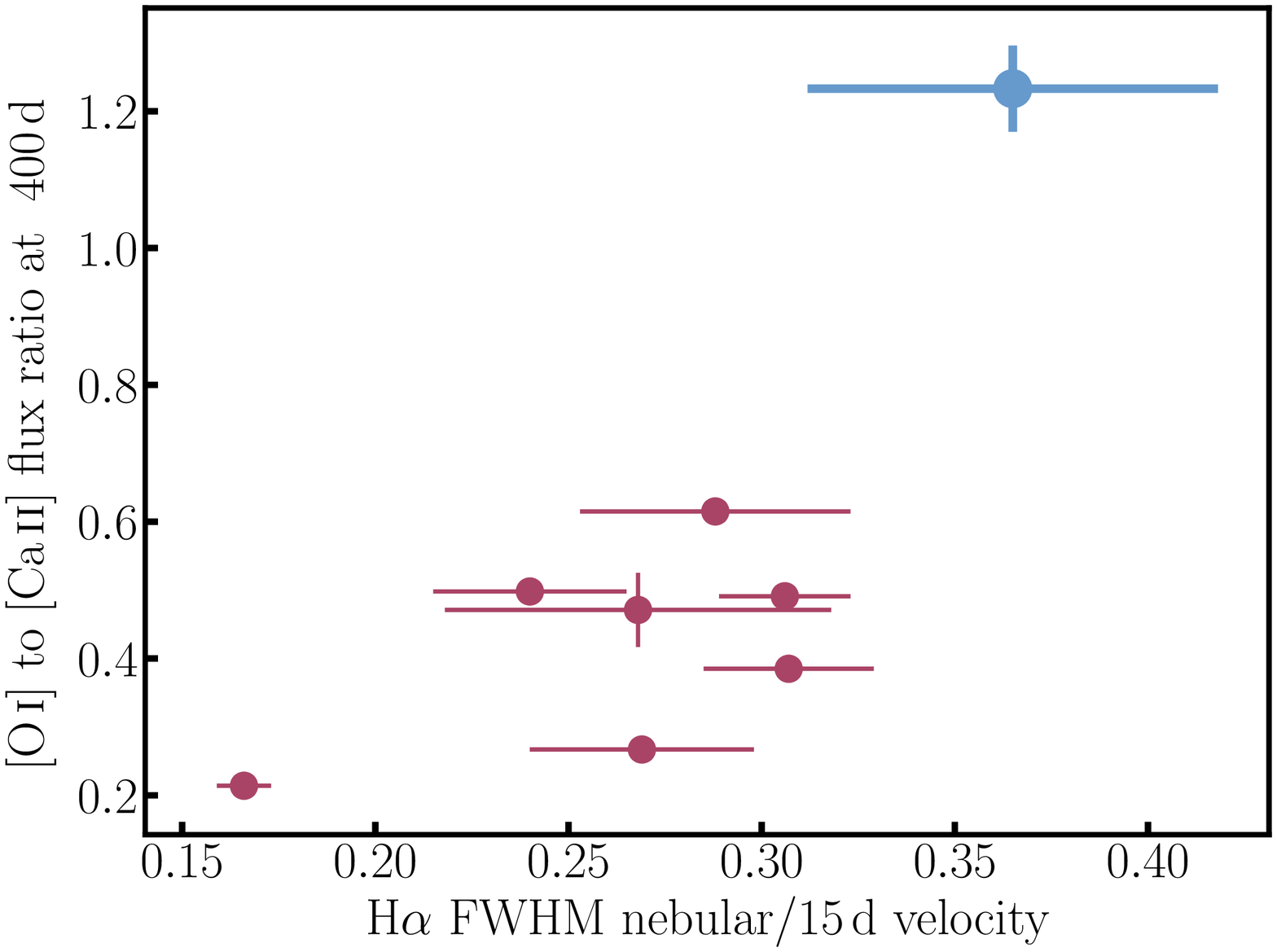}
\caption{Ratio of the SN~II integrated 
nebular [\oi] 6300,6364 \AA\ to [\caii] 7291,7323 \AA\ fluxes plotted against 
the SN~II ratio of the \ha\ FWHM velocity at nebular epochs with respect to the photospheric velocity
+15\,d. SN~2015bs is shown in blue, with 
the rest of the sample in maroon. 
It is clear that in the y-axis SN~2015bs is very much at the extremity of the distribution, while in the x-axis SN~2015bs displays the
highest ratio of nebular to photospheric-phase ejecta velocities. 
Ratio errors are propagated errors from the standard deviation of multiple spectral measurements of each feature with slight changes in the 
defined continuum level each time.
(See Section 3 of Methods for details of how these parameters 
were measured for each SN.)}
\label{rationeb}
\end{suppfigure}

\begin{suppfigure}
\includegraphics[width=17cm]{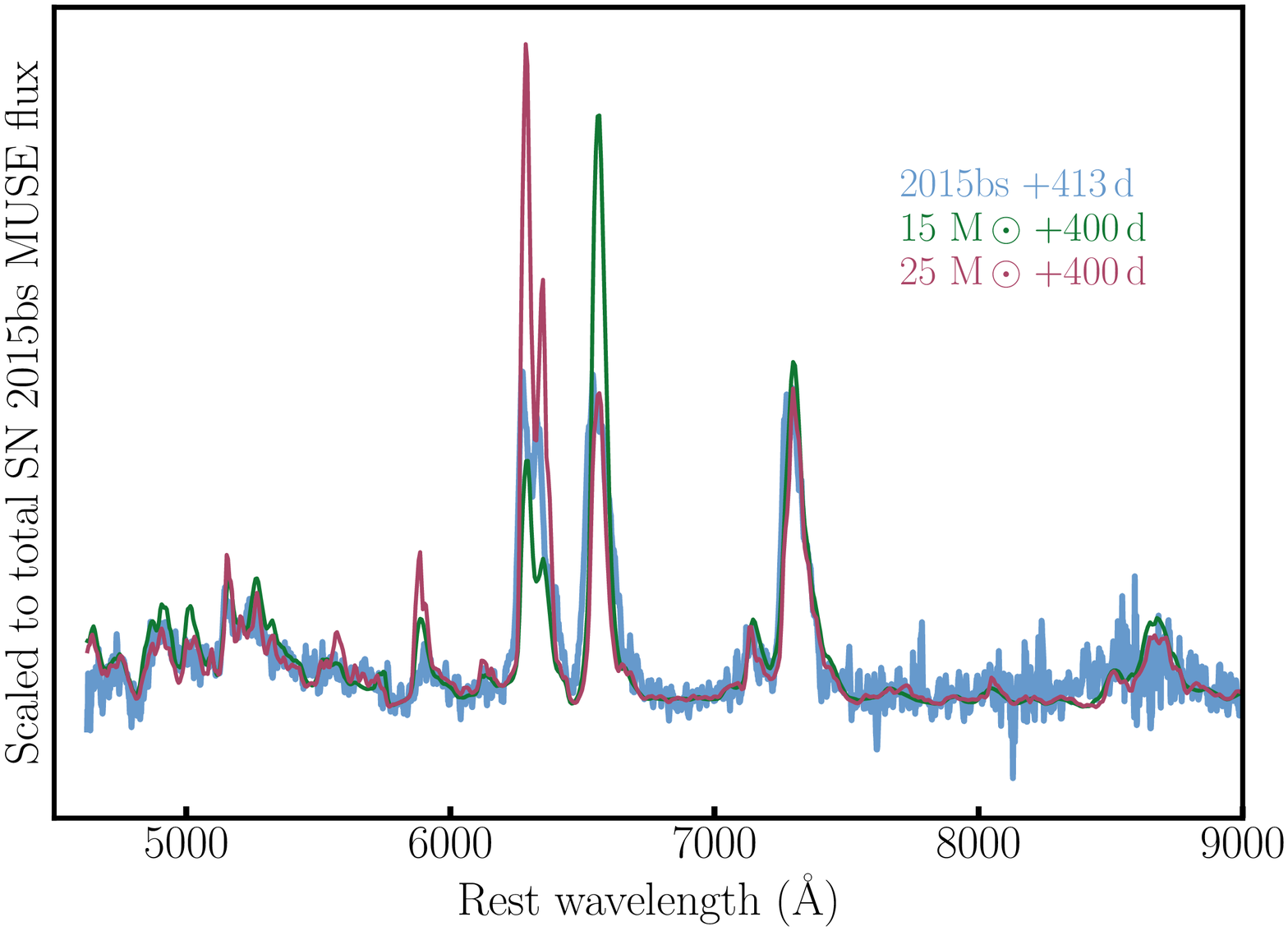}
\caption{Comparison of the nebular spectrum of SN~2015bs with model spectra at similar epochs. The model
spectra are produced by progenitors of 15 and 25\msun\ at solar metallicity\cite{jer14}.
The spectra are scaled
to the total flux contained within the MUSE wavelength range of the SN~2015bs spectrum.}
\label{15bsjerk}
\end{suppfigure}

\begin{suppfigure}
\includegraphics[width=17cm]{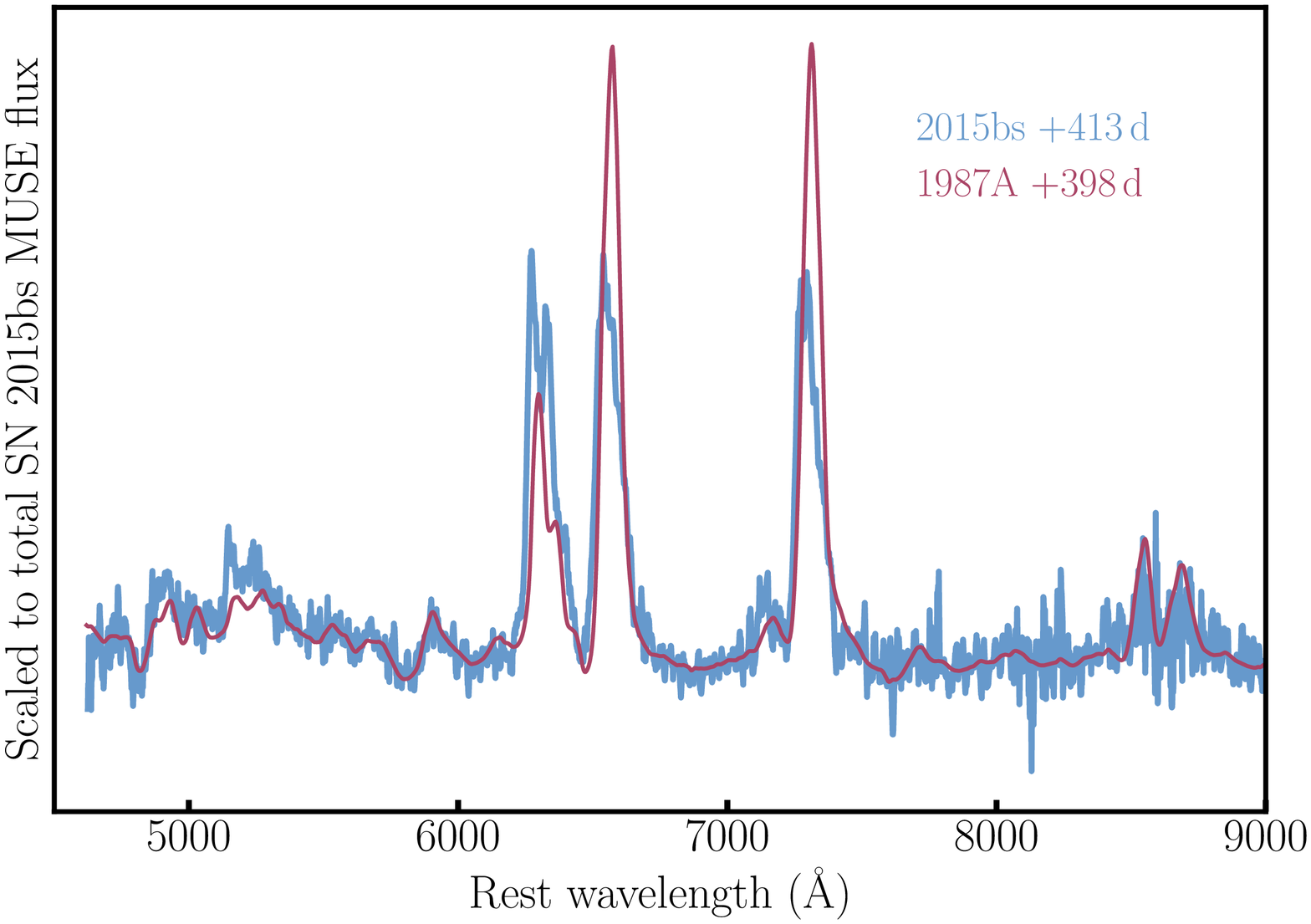}
\caption{Comparison of the SN~2015bs nebular spectrum with that at a similar epoch of SN~1987A. The spectra are scaled
to the overal flux contained within the MUSE wavelength range of the SN~2015bs spectrum. Note the relative strength of
[\oi] in SN~2015bs with respect to SN~1987A, suggesting a higher progenitor mass for the former. In addition, this plot
clearly shows the asymmetric nature of the emission lines of SN~2015bs.}
\label{15bs87a}
\end{suppfigure}

% Your references go at the end of the main text, and before the
% figures.  For this document we've used BibTeX, the .bib file
% scibib.bib, and the .bst file Science.bst.  The package scicite.sty
% was included to format the reference numbers according to *Science*
% style.

\begin{supptable}
\scriptsize
\centering
\caption{SN~2015bs local sequence photometry in the standard systems\cite{lan92,smi02}}
\begin{tabular}[t]{ccccccc}
\hline
Star & RA & Dec & $B$ (mag) & $V$ (mag) & $r$ (mag) & $i$ (mag)\\
\hline
1    & 22:33:23.62  &  $-$6:24:24.31& 13.827 (0.003)&$\cdots$       &$\cdots$       &$\cdots$      \\
2    & 22:33:23.20  &  $-$6:28:33.77& 14.400 (0.009)&13.849 (0.013) &$\cdots$       &$\cdots$      \\
3    & 22:34:09.33  &  $-$6:15:45.02& 17.382 (0.079)&16.569 (0.023) &16.211 (0.032) &15.940 (0.013)\\
4    & 22:33:44.86  &  $-$6:18:27.14& 15.000 (0.015)&14.295 (0.003) &$\cdots$       &$\cdots$      \\
5    & 22:33:53.94  &  $-$6:22:53.67& 15.049 (0.006)&14.424 (0.003) &14.204 (0.003) &$\cdots$      \\
6    & 22:33:49.00  &  $-$6:16:09.63& 15.280 (0.015)&14.529 (0.006) &14.230 (0.006) &$\cdots$      \\
7    & 22:34:03.75  &  $-$6:21:08.24& 15.812 (0.012)&14.898 (0.007) &14.458 (0.006) &14.158 (0.009)\\
8    & 22:33:55.22  &  $-$6:18:58.88& 15.740 (0.010)&14.929 (0.009) &14.541 (0.012) &14.232 (0.021)\\
9    & 22:33:32.02  &  $-$6:18:42.94& 15.920 (0.018)&15.190 (0.004) &14.901 (0.013) &14.645 (0.015)\\
10   & 22:33:26.33  &  $-$6:25:43.01& 16.268 (0.043)&15.291 (0.018) &14.766 (0.015) &14.260 (0.010)\\
11   & 22:34:12.19  &  $-$6:19:20.36& 16.173 (0.038)&15.481 (0.010) &15.212 (0.016) &15.001 (0.014)\\
12   & 22:33:59.20  &  $-$6:18:44.41& 16.134 (0.011)&15.512 (0.010) &15.292 (0.009) &15.119 (0.014)\\
13   & 22:34:07.89  &  $-$6:20:13.86& 16.046 (0.017)&15.535 (0.013) &$\cdots$       &15.232 (0.017)\\
14   & 22:33:33.85  &  $-$6:29:09.08& 16.624 (0.033)&15.782 (0.009) &15.381 (0.012) &15.070 (0.013)\\
15   & 22:33:22.22  &  $-$6:16:26.54& 16.678 (0.047)&16.145 (0.032) &15.947 (0.019) &15.754 (0.014)\\
16   & 22:33:26.96  &  $-$6:23:44.69& 17.077 (0.051)&16.217 (0.020) &15.838 (0.016) &15.524 (0.014)\\
17   & 22:34:03.21  &  $-$6:14:55.80& 17.347 (0.063)&16.275 (0.013) &15.677 (0.023) &15.212 (0.012)\\
18   & 22:33:21.69  &  $-$6:27:28.13& $\cdots$      &16.215 (0.014) &15.840 (0.021) &15.580 (0.022)\\
19   & 22:33:34.72  &  $-$6:25:11.91& 17.368 (0.093)&16.430 (0.019) &15.931 (0.021) &15.567 (0.018)\\
20   & 22:34:13.70  &  $-$6:29:22.95& 17.291 (0.017)&16.461 (0.019) &15.988 (0.023) &15.749 (0.010)\\
21   & 22:33:49.54  &  $-$6:17:59.55& 17.165 (0.011)&16.471 (0.018) &16.175 (0.027) &15.967 (0.027)\\
22   & 22:34:06.06  &  $-$6:24:04.69& 17.006 (0.060)&16.516 (0.033) &16.347 (0.024) &16.200 (0.024)\\
23   & 22:34:02.12  &  $-$6:24:44.69& 17.331 (0.050)&16.551 (0.020) &16.223 (0.030) &15.914 (0.028)\\
24   & 22:33:40.60  &  $-$6:28:32.55& 17.272 (0.066)&16.573 (0.004) &16.246 (0.007) &15.970 (0.035)\\
\hline	
\hline
\end{tabular}
\setcounter{supptable}{0}
\caption{\footnotesize In the first column local sequence star number is given. 
In columns 2 and 3 the RA and Dec coordinates are listed for each star (J2000.0).
In columns 4, 5, 6, and 7 magnitudes are listed for each star as observed in the $B-$, $V-$, $r-$ and $i-$band respectively.
Magnitude errors are given in parenthesis. Photometric errors are the propagated errors from the photometric calibration.}
\label{tablocal}
\end{supptable}

\newpage

\begin{supptable}
\scriptsize
\centering
\caption{SN~2015bs photometry in the natural system of the Swope telescope}
\begin{tabular}[t]{cccccc}
\hline
JD  & Epoch (days post explosion) & $B$ (mag) & $V$ (mag) & $r$ (mag) & $i$ (mag)\\
\hline
2456938.6 & 18 & 18.249 (0.034) & 18.032 (0.029) & 17.880 (0.025) & 18.015 (0.027)\\
2456939.6 & 19 & 18.332 (0.032) & 18.015 (0.023) & 17.860 (0.019) & 18.017 (0.022)\\
2456940.5 & 20 & 18.288 (0.019) & 18.051 (0.019) & 17.854 (0.017) & 17.997 (0.018)\\
2456941.5 & 21 & 18.367 (0.015) & 18.042 (0.013) & 17.899 (0.014) & 17.992 (0.017)\\
2456942.5 & 22 & 18.380 (0.010) & 18.067 (0.012) & 17.884 (0.011) & 17.995 (0.013)\\
2456944.5 & 24 & 18.459 (0.014) & 18.058 (0.016) & 17.903 (0.014) & 18.004 (0.017)\\
2456947.6 & 27 & 18.559 (0.011) & 18.123 (0.013) & 17.858 (0.078) & $\cdots$      \\       
2456948.5 & 28 & 18.575 (0.012) & 18.123 (0.012) & 17.932 (0.011) & 18.042 (0.016)\\
2456949.5 & 29 & 18.602 (0.012) & 18.133 (0.011) & 17.918 (0.011) & 18.036 (0.013)\\
2456956.5 & 36 & 18.796 (0.016) & 18.229 (0.013) & 17.965 (0.016) & 18.068 (0.016)\\
2456979.5 & 59 & 19.141 (0.018) & 18.323 (0.014) & 18.018 (0.014) & 18.070 (0.017)\\
2456987.5 & 67 & 19.209 (0.019) & 18.352 (0.014) & 18.031 (0.013) & 18.064 (0.020)\\
2456998.5 & 78 & 19.342 (0.068) & 18.519 (0.036) & 18.110 (0.020) & 18.130 (0.023)\\
2456998.5 & 78 & 19.575 (0.083) & 18.484 (0.035) & 18.163 (0.023) & 18.100 (0.021)\\
2457005.5 & 85 & 19.687 (0.047) & 18.667 (0.030) & 18.272 (0.042) & 18.256 (0.039)\\
2457013.5 & 92 & 20.173 (0.052) & 18.996 (0.027) & 18.558 (0.019) & 18.430 (0.024)\\
\hline	                                 
\hline      
\end{tabular}
\setcounter{supptable}{1}
\caption{\footnotesize In the first column the Julian Date of the observations is listed, followed by the
epoch post explosion in column 2. The $B-$, $V-$, $r-$, and $i-$band photometry are
then
listed in columns 3, 4, 5, and 6 respectively.
Magnitude errors are given in parenthesis. Photometric errors are the propagated errors from the photometric calibration.}
\label{tabphot}
\end{supptable}

\begin{supptable}
\scriptsize
\centering
\caption{SN~2015bs $w_{\rm ps}$-band photometry from Pan-STARRS}
\begin{tabular}[t]{cccccc}
\hline
MJD  & Epoch (days post explosion) &  $w_{\rm ps}$ (mag)\\
\hline
57200.5 & 280 &21.38 (0.08)\\ 
57200.5 & 280 &21.46 (0.07)\\
57200.5 & 280 &21.52 (0.10)\\
57200.6 & 280 &21.40 (0.08)\\
57226.4 & 306 &21.73 (0.19)\\
57243.4 & 323 &21.88 (0.21)\\
57275.3 & 355 &22.33 (0.20)\\ 
57277.3 & 357 &22.31 (0.17)\\
\hline	                                 
\hline      
\end{tabular}
\setcounter{supptable}{2}
\caption{\footnotesize In the first column the Modified Julian Date of the observations is listed, followed by the
epoch post explosion in column 2. Photometry is listed in columns 3.
Magnitude errors are given in parenthesis. Photometric errors are the propagated errors from the photometric calibration combined
with those arising from the difference imaging.}
\label{tabpan}
\end{supptable}

\begin{supptable}
\scriptsize
\centering
\caption{SN~II nebular comparison sample}
\begin{tabular}[t]{cccc}
\hline
SN & Photometry & Photospheric-phase spectroscopy & Nebular spectroscopy\\
\hline
1999em & Galbany et al. (2016)\cite{gal16}   & unpublished                        &  Elmhamdi et al. (2003)\cite{elm03}\\
2004et & Misra et al. (2007)\cite{mis07}  & Sahu et al. (2006)\cite{sah06}  &  Maguire et al. (2010)\cite{mag10}\\
2007aa & unpublished                         & unpublished                        &  Maguire et al. (2012)\cite{mag12}\\
2009N  & Takats et al. (2014)\cite{tak14}  & Takats et al. (2014)\cite{tak14}   &  Maguire et al. (2012)\cite{mag12}\\
2012A  & Tomasella et al. (2013)\cite{tom13} & Tomasella et al. (2013)\cite{tom13}&  Tomasella et al. (2013)\cite{tom13}\\
2012aw & Bose et al. (2013)\cite{bos13}     & Dall'Ora et al. (2014)\cite{dal14} &  Jerkstrand et al. (2014)\cite{jer14}\\
2013ej & Bose et al. (2015)\cite{bos15}      & Yuan et al. (2016)\cite{yua16}   &  Yuan et al. (2016)\cite{yua16}\\
\hline
\hline
\end{tabular}
\setcounter{supptable}{3}
\caption{\footnotesize References for data from the comparison SN~II sample that have nebular spectrscopy within $\pm$50 days
of that of SN~2015bs. In the first column the SN name is listed. In columns 2, 3, and 4 we list the reference for the photometry,
photospheric-phase spectroscopy, and nebular-phase spectroscopy respectively.}
\label{tabcomp}
\end{supptable}

\begin{supptable}
\scriptsize
\centering
\caption{SN~2015bs measured parameters}
\begin{tabular}[t]{ccccccccc}
\hline
& $M_{\rm max}$ (mag) & $M_{\rm end}$ (mag) & $s_{\rm 1}$ (mag 100d$^{-1}$) & $s_{\rm 2}$ (mag 100d$^{-1}$)& ${\rm Pd}$ (days)& 
${\rm OPTd}$ (days) & $^{56}{\rm Ni}$ (\msun)\\
\hline
SN~2015bs            & --17.54$\pm$0.05 & --17.19$\pm$0.05 & 1.17$\pm$0.10 & 0.33$\pm$0.04 & 39.5$\pm$4 & 78.5$\pm$5 & 0.048$\pm$0.008\\
Mean A14 &--16.74$\pm$1.01 & --16.03$\pm$0.81 & 2.65$\pm$1.50 & 1.27$\pm$0.93 & 48.4$\pm$12.6 & 83.7$\pm$16.7 & 0.032$\pm$0.021\\
\hline
\end{tabular}
\setcounter{supptable}{4}
\caption{\footnotesize Measured parameters for SN~2015bs in the first row, and mean values from a large sample\cite{and14} in the second row.
In column 2 we list $M_{\rm max}$, the absolute $V$-band magnitude at maximum light. In column 3 $M_{\rm end}$ , the absolute $V$-band magnitude
at the end of the plateau. $s_{\rm 1}$ is listed in column 4 and $s_{\rm 2}$ in column 5, corresponding to the initial decline
rate from maximum, and the second shallower decline rate during the plateau respectively (both in the $V$ band).
In column 6 we list ${\rm Pd}$, the plateau duration, which is defined from the inflection point between $s_{\rm 1}$ and $s_{\rm 2}$ 
until the end of the plateau. ${\rm OPTd}$, the optically thick phase duration, is listed in column 7 and is the time duration
between explosion and the end of the plateau. Finally in column 8 we give the $^{56}{\rm Ni}$ mass.
Errors on the mean values are the standard deviation of the each distribution. Errors on $M_{\rm max}$ and $M_{\rm end}$
for SN~2015bs come from the uncertainties on the SN photometry combined with the distance uncertainty, 
while errors on the decline-rate measurements $s_{\rm 1}$ and $s_{\rm 2}$
are the straight-line fit errors on each parameter. The error on ${\rm OPTd}$ is dominated by the error on the explosion epoch of SN~2015bs, and
the error on ${\rm Pd}$ arises from a combination of the error on the $s_{\rm 1}$--$s_{\rm 2}$ transition epoch and the error on the definition 
of the end of the plateau. Finally, the error on the $^{56}{\rm Ni}$ mass is the standard deviation of the three independent measurements are outlined
in the text.}
\label{tabpar}
\end{supptable}

%% Here is the endmatter stuff: Supplementary Info, etc.
%% Use \item's to separate, default label is "Acknowledgements"

%%
%% TABLES
%%
%% If there are any tables, put them here.
%%

\end{document}